\documentclass[preprint,12pt]{elsarticle}

\usepackage{amssymb}
\usepackage[dvipdfm]{hyperref}

\journal{Astroparticle Physics}

\begin{document}
\newcommand{\el}{\mbox{${\rm e^{-}}$ }}
\newcommand{\ps}{\mbox{${\rm e^{+}}$ }}

\begin{frontmatter}



\title{Chemical Composition of Galactic Cosmic Rays with Space Experiments}


\author[1]{Mirko Boezio}
\author{Emiliano Mocchiutti}

\address{INFN, Sezione di Trieste, I-34149
Trieste, Italy}

\begin{abstract}
The origin and properties of the cosmic radiation are one of the most 
intriguing question in modern astrophysics. The precise measurement of the 
chemical composition and energy spectra of the cosmic rays provides 
fundamental insight into these subjects. In this paper we will review the 
existing experimental data. Specifically, we will analyse results 
collected by space-born experiments discussing the experimental 
uncertainties and challenges with a focus on the PAMELA experiment.
 
\end{abstract}


\end{frontmatter}



\section{Introduction}
\label{intro}

Since the discovery of cosmic rays, the fundamental question concerned
their origin, acceleration and propagation mechanisms.  From an energetic
point of view it was realized in the fifties
(e.g.~\cite{Gin64}) that Super Nova
explosions, already considered as cosmic-ray sources by Baade and
Zwicky~\cite{bad34} in the thirties,  
released sufficient energy to power the cosmic rays in the
Galaxy.
A viable acceleration 
mechanism, based on diffusive shock acceleration (``first order
Fermi mechanism'') 
produced by supernova (SN) shock waves propagating in the interstellar medium 
(see \cite{mal01} for a review), 
was proposed in the seventies and recent
measurements of synchrotron X-rays and TeV gamma-rays point
unambiguously to supernovae remnants (SNR) \cite{all97,aha04}
as the acceleration site of at least cosmic-ray electrons.
However, the
vast majority of the cosmic rays are protons and heavier nuclei with
electrons representing only few percent of the total flux.
For example, TeV emission from the young supernova
remnant RX J1713.7-3946, detected by the
H.E.S.S. collaboration~\cite{aha07}, has been interpreted as
originating from hadronic interactions of cosmic rays with energies
above  $10^{14}$ eV  in the 
shell of the SNR (even though leptonic processes cannot be ruled
out~\cite{por06,ell10}). X-ray
measurements of the same SNR provide evidence that protons and nuclei
can be accelerated to energies  $\geq 10^{15}$ eV
\cite{uch07}. 
Recent AGILE observations of diffuse gamma ray emission in the 100 MeV
- 1 GeV range from the outer shock region of SNR IC 443 have been
explained in terms of hadronic
acceleration~\cite{tav10}. Likewise, Fermi observations of the
shell of SNR W44 have been attributed to the decay of  
$\pi^0$s produced during interactions of accelerated hadrons with the
interstellar medium~\cite{abd10}. 

At the end of the acceleration phase, particles are injected into the
interstellar medium where 
they propagate, diffusing in the turbulent galactic magnetic
fields. Nowadays, this propagation 
is well described by solving numerically (e.g. the  GALPROP simulation
code \cite{str98}) or analytically
(e.g.~\cite{jon01,don01}) the transport equations
for the particle diffusion in the galaxy. 
The galactic magnetic field masks the arrival direction of charged
particles, making the flux 
isotropic. Hints of galactic anisotropy have been reported in the
multi-TeV region \cite{ame06,abd09b,abb09,det11}.  

One of the features predicted by such acceleration and propagation
models is that the cosmic-ray spectra are well described by
single power laws, with 
similar spectral indices ($\gamma \simeq -2.7$) for protons and
heavier nuclei, up to energies of  $\approx 10^{15}$ eV 
(the so called `knee' region). Till recently, indeed, observations
looked consistent with such picture.
More recent acceleration models account for the dynamical interaction
between the shock front and accelerated particles. The resulting
energy injection spectra are not anymore a single power law 
but are in general concave presenting a hardening 
at higher energies (e.g. \cite{cap10}). Combined with propagation
effects, additional structures may 
appear in the energy spectra of cosmic rays probably in the TeV region
with possible spectral differences between the various species
(e.g. \cite{bla11}).

About 50 years ago a clear change in the energy
spectrum of cosmic rays was observed around $10^{15}$ eV~\cite{kul58}. Since then, more
data have been acquired confirming the first evidence, but no
theoretical explanation have yet been accepted as fully satisfactory
by the cosmic-ray community. While yet unclear, the origin of the knee
is probably related to the acceleration mechanism. In fact in
diffusive shock acceleration model cosmic rays are accelerated in
blast waves of SNR and a rigidity (R = pc/(Ze),
p being the momentum of a particle of charge Ze)-dependent limit, above which the
diffusive shock acceleration becomes inefficient, is predicted. The
maximum energy attainable by a nucleus of charge Z may range from
Z$ \times 10^{14}$ eV to Z$\times 10^{15}$ eV depending on the model
and types of supernovae 
considered~\cite{lag83}. 
Then, the knee would result from the convolution of the
various cutoffs while the spectral composition would become heavier. 
An alternative explanation of the knee is adopted by models that
relate it to leakage of cosmic rays from the Galaxy. In this case the
knee is expected to occur at lower energies for light nuclei as
compared to heavy ones, due to the rigidity-dependence of the Larmor
radius of cosmic rays propagating in the galactic magnetic
field~\cite{hor04}.
The majority of the data in the knee region have
been 
collected by ground detector arrays that, measuring the secondary
particles produced by cosmic rays interacting with the Earth's
atmosphere, indirectly determine the energy and composition of the
cosmic radiation. At lower energies, up to about $10^{14}$ eV the 
cosmic-ray spectra have been directly measured mostly by balloon-borne
experiments. Both the statistical and systematic
uncertainties of these measurements significantly hinder the
interpretation of the data. 

The cosmic-ray spectra observed at Earth result from the combined
effects of acceleration and propagation. A powerful test of
propagation models is the measurement of secondary nuclei (e.g. Boron)
that are not end-points of stellar evolution but are produced by the
interaction of primary cosmic rays (e.g. Carbon and Oxygen) with the
interstellar matter. The comparison between their energy spectra and
their parent energy spectra (e.g. the B to C ratio, see Fig. \ref{fig:bc} on the left, or the
ratio of sub-iron elements to iron) provides fundamental information
about the secondary production and propagation of cosmic rays and
their dependence from the cosmic-ray energy (or rigidity). 

While
convincing, the evidences of SNR as acceleration sites of all galactic cosmic
rays are not conclusive and this is especially true for the
acceleration mechanisms proposed to explain the cosmic-ray
spectrum. Moreover, the paucity of high energy (TeV) data concerning
secondary nuclei 
seriously limits our understanding of the
interplay between propagation and acceleration. 
Furthermore, the
recent observations by PAMELA~\cite{adr11a}, CREAM~\cite{ahn10} and ARGO-YBJ~\cite{det11}
point to spectral
shapes depending on the nuclei species and deviating from the single
power law dependence.

Precise determination of
the cosmic-ray fluxes and compositions 
is of crucial importance for the understanding of astrophysical
phenomena taking place in the Galaxy. Moreover, data from space
mission are sorely needed since the accurate determination of the
chemical composition for balloon-borne experiments is 
intrinsically limited to the region below a few hundred GeV per
nucleon because of uncertainties in the atmospheric corrections
(i.e.. secondaries produced by cosmic rays interacting with the
residual atmospheric overburden). 

Here we will review the existing data on the chemical
composition of cosmic rays 
from space-borne
experiments, with a special focus on the data produced by the
PAMELA experiment~\cite{pic07}. We will analyse the experimental uncertainties
and challenges and provide a brief outlook for the future of this
experimental field.

\section{Proton and Helium data}
\label{s:PHe}

\begin{figure}[!ht]
\includegraphics[width=25pc]{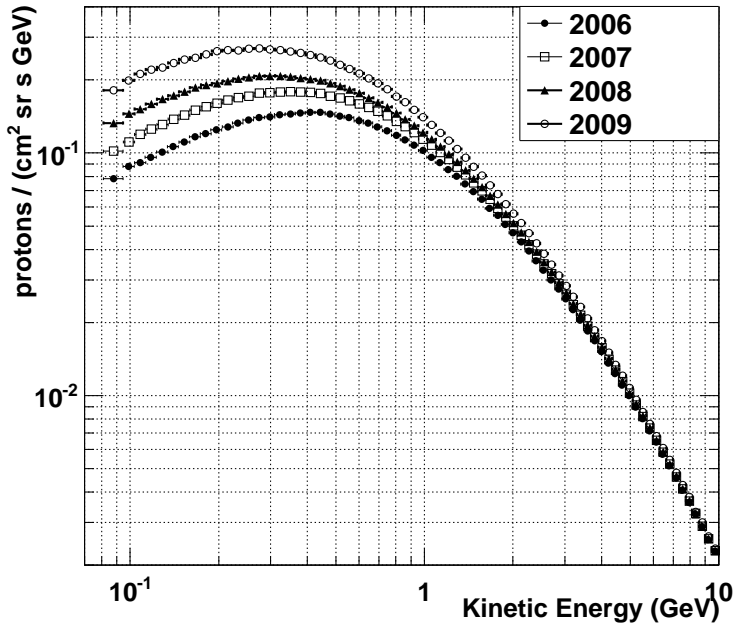}\hspace{2pc}
\caption{The yearly proton energy spectrum measured by PAMELA from the
  beginning of the space mission in mid of 2006 till end of 2009~\cite{des11}. 
\label{fig:Pmod}}   
\end{figure}

Measurements of primary cosmic-ray proton and helium nuclei spectra
have been performed over the years using different
techniques: 
magnet spectrometers, e.g. \cite{boe99}, and RICH detectors
\cite{buc94} have been used for energies up to 1 TeV/n, while
calorimetry measurements 
extended to higher energies, e.g. \cite{asa98}. 
The majority of these results, especially concerning 
the high-energy ($> 1$ GeV) part of the spectra ,
were obtained by balloon-borne
experiments. 
Primary cosmic-ray data from space-borne experiments refer
mostly to energies lower than 1~GeV. For example, the series of IMP
(Interplanetary Monitoring Platform) spacecrafts and especially the 
University of Chicago's Cosmic Ray Nuclear Composition experiment aboard the 
IMP-8 satellite \cite{gar75} measured the energy
spectra and chemical composition of cosmic rays up to $\sim$ 100
MeV. While providing relevant information concerning the inner
heliospheric condition and effects of solar activity on the cosmic
radiation, the data are of less immediate use for studies of galactic
cosmic rays. In fact, the solar wind 
significantly affects the low energy part of cosmic rays, as
it can be easily verified monitoring the time variation of the
cosmic-ray fluxes and their dependence with the sun activity. 
As an example Fig.~\ref{fig:Pmod}
shows the yearly low energy proton flux measured by PAMELA from July
2006 till December 2009, i.e. during the last extended solar
minimum. The flux variation with time can be clearly seen as well as 
the decreasing significance of the variation as the energy increases,
becoming negligible above 5~GeV.

The
interplay between solar wind and cosmic radiation in the heliosphere is
a relevant and fundamental problem in space plasma physics,
heliospheric physics and in cosmic ray physics. This is a very active
field complementing state-of-the-art numerical models (e.g.~\cite{pot09})
with a
significant amount of new data. 
The heliosphere is the only astrophysical system which is accessible
to in-situ spacecraft measurements and its modeling can also lead to
fundamental insights applicable to larger astrophysical systems. When
the transport and modulation of cosmic rays in the heliosphere will be
fully understood it will be possible to infer the local interstellar
spectra of cosmic rays down to very low energies (tens of MeV) thus
enabling significant conclusions on cosmic ray acceleration and
propagation mechanisms (e.g. \cite{del10}). However, the existing
theoretical 
and experimental data are not yet sufficiently precise to disentangle the
galactic and heliospheric effects on the low energy cosmic
radiation. On the other hand, the solar effects on the cosmic-ray
energy spectra above several GeV are sufficiently low 
(Fig.~\ref{fig:Pmod}) that these data
can be considered unbiased samples of the local interstellar
spectra. 

\begin{figure}[!ht]
\begin{center}
\includegraphics[width=25pc]{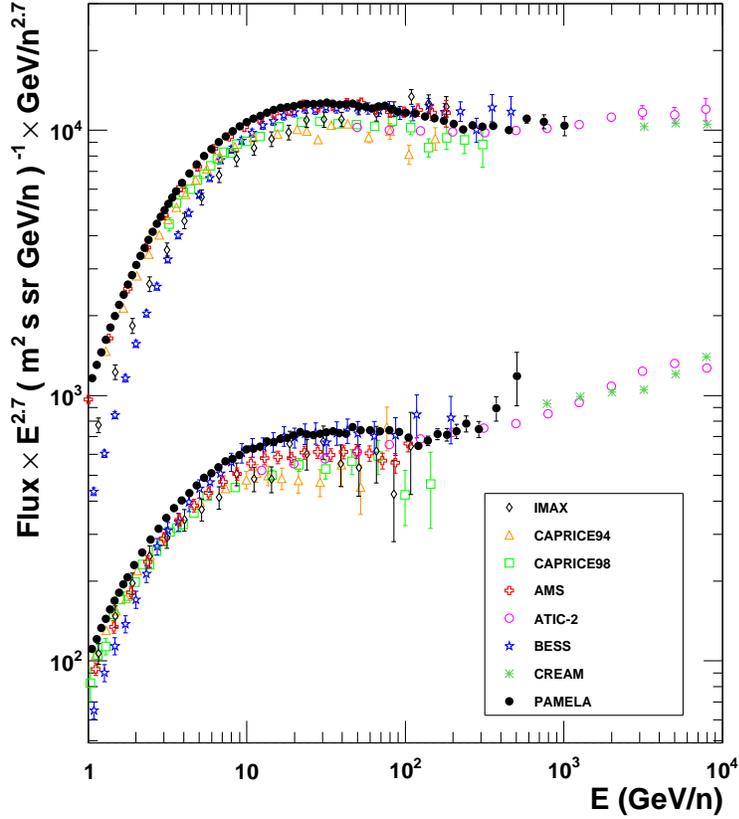}\hspace{2pc}
\end{center}
\caption{Recent results on the proton and helium energy spectra 
above 1 GeV/n obtained by balloon-borne: 
CAPRICE94 \cite{boe99}, IMAX \cite{men00}, CAPRICE98
\cite{boe03}, BESS \cite{hai04}, ATIC-2 \cite{wef08},
CREAM \cite{ahn10}, 
and
space-borne: AMS-01 \cite{alc00a}, PAMELA \cite{adr11a}
experiments.}
\label{fig:PHeflux}
\end{figure}

Figure~\ref{fig:PHeflux} shows the proton and helium energy 
spectra\footnote{As usually done, the 
  fluxes are multiplied by E$^{2.7}$, where E is the energy in GeV. Reducing
  the decades of variation of the flux, this
  allows for a clearer picture of the spectral shapes. However,
  this implies that the absolute energy uncertainties are added to the
  flux uncertainties.} above 1~GeV/n 
measured by recent
balloon- \cite{ahn10,boe99,men00,boe03,hai04,wef08}
and
space-borne \cite{adr11a,alc00a}
experiments. 
Several conclusions can be drawn from these 
data:
\begin{itemize}
  \item the high energy (above 1 TeV/n) measurements can be
    reconciled with the lower energy data only assuming a
    hardening of the spectra in the hundred GeV region as explicitly
    indicated by
    PAMELA and ATIC-2 results;
  \item the most recent data from PAMELA~\cite{adr11a} and CREAM~\cite{ahn10} show
    spectral 
    differences between the proton and helium nuclei.
  \item all sets of results are in relatively good agreement, but
    differences well beyond the quoted statistical-only errors can be
    noticed at high energies\footnote{Below 10 GeV, the difference
      between the various results is mainly due to solar modulation
      effects, since the experiments were performed at different epochs
      of solar activity.}; 
\end{itemize}
The first two results are interesting and compel a revision of the
paradigm of cosmic ray acceleration in supernova remnants followed by
diffusive propagation 
in the galaxy. 
The discussion about the possible interpretations
is beyond the scope of this work. However, as a short
summary, these results may indicate that the SNR paradigm has to be
better 
understood as, for example, including the
stochasticity in the spatial and temporal distribution of
SNR~\cite{bla11}. Or different populations of cosmic-ray sources
should be considered such as, for example, novae stars and explosions
in superbubbles \cite{zat06}, 
different acceleration sources for protons and helium and heavier
nuclei \cite{bie93}.

Here we will discuss the experimental challenges and
significances of these
measurements in light of the last two points presented above.

\subsection{Systematic Uncertainties}
Proton and helium nuclei are usually experimentally identified by measuring their 
ionization losses in charge sensitive detectors such as plastic
scintillators or silicon-array detectors. Since they 
are the 
two most abundant components in the cosmic radiation, a charge
selection is sufficient to provide clean proton and helium
nuclei samples with negligible contamination of other particles
(positrons for proton measurements in space, muons for proton
measurements at balloon altitude, accounting for a percent or less of
high-energy protons). The only significant contamination is of protons
in the helium sample and viceversa. This can be reduced to a negligible
amount by using redundant ionization measurements. Therefore, the
reason for the differences in the experimental measurements, as seen
in Fig.~\ref{fig:PHeflux}, cannot be due to background issues.
 
For the balloon measurements, an additional systematic uncertainty is
related to the correction for the residual atmospheric overburden,
both as production of secondaries and losses due to
interaction. However, this correction 
usually amounts to less than 10\% with a consequent per-cent 
uncertainty on the estimated flux. 

Quite likely 
the main sources of 
discrepancy arise from 
efficiency and energy determinations. 
Selection efficiencies are an experimental challenge since they
require a very good knowledge of the detector performances during
data taking. Often, to reduce the systematic uncertainties, the
selection efficiencies are derived from flight data, but a fully
unbiased cross calibration of the efficiencies in flight is quite
impossible and simulations have to be used. The simulations are
validated by comparisons with test-beam data, which do not
account for the flight condition, and, whenever possible, flight
data. However, unproven assumptions have to be made resulting in 
uncertainties that have to be included in the results.
It has to be noted that efficiency uncertainty usually affects the
absolute normalization of the fluxes and have a smaller impact on the
shape of the spectra. 

Another major possible source of discrepancies between the
measurements is the energy determination. In the two satellite
experiments (AMS-01 and PAMELA) the energy, or more precisely the
rigidity, was determined measuring
the curvature of charged particles in a magnetic spectrometer. In both
cases the particle tracks were reconstructed interpolating position
points measured in silicon tracking devices inserted in the magnetic
field of permanent magnets. The relevant quantity for this measurement
is the 
Maximum Detectable Rigidity (MDR) defined
as the rigidity for which the relative error on the rigidity $\Delta
R/R=100\%$. 
The momentum resolution and MDR of the magnetic spectrometer  depend on
the spatial resolution in the bending view and on the topology of the
event.  

\paragraph{The PAMELA case}
For each event the track fitting procedure determined the deflection
($\pm 1/R$) of the particle. The error associated with the measured deflection was used as
an estimate 
of the MDR for each  event. 
This MDR varied from 200 GV to about 1.5 TV according to the distance 
between the two most distant silicon tracking layers with a detected
signal 
of the reconstructed track. The presented results
(Fig.~\ref{fig:PHeflux}) were 
obtained using events for which the measured rigidity was smaller than
the estimated MDR (hence: R $<$ MDR). Consequently, the
reconstructed energy spectra had to be unfolded for the varying
and increasing, as the rigidity increases, errors on the event
deflections. The 
unfolding procedure used a standard Bayesian approach as described in
\cite{dag95}. 
This procedure relied on a simulation of the apparatus,
which was validated by comparing the distributions of several
significant variables (e.g. coordinate residuals, chi-square and
the covariance matrix from the track fitting) with those obtained from
real data. 
A systematic uncertainty, estimated
folding and unfolding a known spectral shape with the spectrometer
response, of 2\% was added to the data. 
Since a clear break in the experimental spectra 
is found around 
200~GV, one may 
wonder if it results from an incomplete unfolding of the spectra. 
An important ingredient for a proper definition of the unfolding
procedure is 
the tracking
alignment. In fact, a wrong assumption on the absolute position of the
tracking sensor respect to the magnetic field
results in a wrong measurement of the deflection,
essentially appearing as an offset in this measurement. 
In experiments like PAMELA and AMS precise
alignments of the tracking system (e.g. \cite{alc99})
were performed at particle test beam facility prior the launch.
However, the alignment procedure had to be repeated to account for
possible changes in the 
flight configuration 
due to the significant shocks and
vibrations the apparatus underwent during the  operations of launch
and placement in 
orbit. Furthermore, all experiments could be affected by a 
wrongly mapped magnetic field (especially important for strongly
inhomogeneous magnetic fields, e.g.~\cite{boe99}). 
One possible way to constrain this effect is a redundant energy
information. For example, the CAPRICE98
collaboration compared the particle rigidity
reconstructed by a magnetic spectrometer with the particle velocity
derived by measurements of Cherenkov angles obtained with a gaseous
RICH detector~\cite{boe03}. 
In the PAMELA experiment no such comparison was available  
for protons and helium
nuclei but it was possible for electrons and positrons using the
electromagnetic calorimeter. The 16 radiation length PAMELA
calorimeter~\cite{boe02} was designed to 
sample the total 
energy deposited by electromagnetic showers, hence for these particles
it provided a
systematically independent energy estimation 
respect to the rigidity measurement. 
Furthermore, a deflection offset would act oppositely for electrons
and positrons resulting, in case of a positive shift in deflection, 
in an overestimation (underestimation) 
of the rigidity for electrons (positrons). On the other hand, the
energy measured by the electromagnetic calorimeter is insensitive
respect to the charge sign of electrons. Therefore, 
the comparison between energy measured by the calorimeter and rigidity
obtained by the tracking system for both electrons and positrons 
provided a constrain for global distortions of the
tracking system. These
would mimic a track curvature and result in a deflection
offset, applicable to all particles species. The previous method
applied to PAMELA data limited this offset to $10^{-4}$
GV$^{-1}$. The corresponding uncertainty was included in the published
results~\cite{adr11a}.

\begin{figure}[!ht]
\begin{center}
\includegraphics[width=25pc]{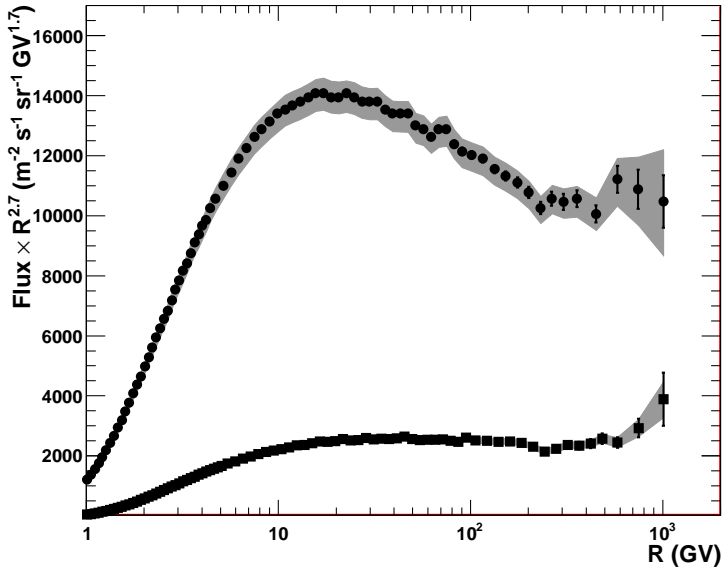}\hspace{2pc}
\end{center}
\caption{Proton (full circles) and helium (full boxes) fluxes measured
  by PAMELA \cite{adr11a}. The 
shaded areas represent the overall (flux and rigidity) estimated
systematic uncertainty.}  
\label{fig:PHeflux2}
\end{figure}

Figure~\ref{fig:PHeflux2}
shows the proton and helium rigidity spectra measured by
PA\-ME\-LA \cite{adr11a}. The 
shaded areas represent the overall (flux and rigidity) estimated systematic uncertainty. For
protons these uncertainties dominate the statistical ones at high
energy as expected from the previous discussion. While large, however,
they cannot explain the difference with the helium spectrum measured
by AMS-01 \cite{alc00a} and other balloon
experiments \cite{boe99,men00,boe03} as shown in
Fig.~\ref{fig:PHeflux}. However, it has to be noted, that the
experimental errors in Fig.~\ref{fig:PHeflux} are only statistical,
hence the residual difference may be due to systematic 
uncertainties of the other experiments. For example,
CAPRICE94 \cite{boe99} quoted a systematic 
uncertainty of 10\% at high energies. 
  
Since they are comparable to the statistical errors,
systematic uncertainties have to be included in any interpretation of the
experimental data. In the case of  
PAMELA proton results, the hardening of the
spectra around 200 GV has only a 95\% confidence level significance
when systematic errors 
are included \cite{adr11a}, instead of a 99.7\% CL if only statistical
errors are considered. Furthermore, there is no clear recipe for
including 
systematic errors with statistical ones. The
PAMELA collaboration decided to quadratically sum the systematic
uncertainties, in the fair assumption of independence of the errors.
Then, it studied the case in which they were added or subtracted to the
experimental data, i.e. assuming that they acted uniformly in one or
the other direction like maximum errors. 
More significant is the PAMELA proton to helium flux ratio since, when
expressed as a function of rigidity, various systematic uncertainties
related to the track measurements cancel out. In this case the
hypothesis of a constant value for this ratio is incompatible with the data above 10 GV at a level of about 9 standard deviations.

Another effect, albeit less significant,  should be considered when
comparing experimental data 
to theoretical calculations: the experiments cannot separate, except
for small energy ranges, the isotopes. Therefore, what is commonly
called proton spectrum is in reality a hydrogen one. Furthermore, for
measurements with magnetic spectrometer the conversion from rigidity
to kinetic energy is performed assuming pure proton and
$^{4}He$ samples, hence  
neglecting any contribution from 
less abundant deuterium ($^{2}H/^{1}H \simeq 1 \% $) and $^{3}He$
($^{3}He/^{4}He \simeq 10\%$).

\subsection{Proton and helium isotopes}
Hydrogen and helium isotopes in cosmic rays are generally
believed to be of secondary origin, resulting mainly
from the nuclear interactions of primary cosmic-ray $^{4}He$ with the
interstellar medium. 
Among several secondary components the
specific feature of light secondaries such as $^{2}H$ and $^{3}He$ is that
their interaction mean free path is considerably larger than
the escape mean free path for cosmic rays from the Galaxy.
This is not the case for the heavier secondaries, where the escape
mean free path is of the same order or greater than their
interaction length. As a consequence, light secondaries provide
information concerning cosmic-ray interstellar propagation
that is complementary to that obtained from the study of
heavy secondaries and their precise measurement 
could tell if the helium nuclei have the same propagation history as
heavier nuclei (e.g.~\cite{web97}).

\begin{figure}[!ht]
\includegraphics[width=7.5cm,height=7.5cm]{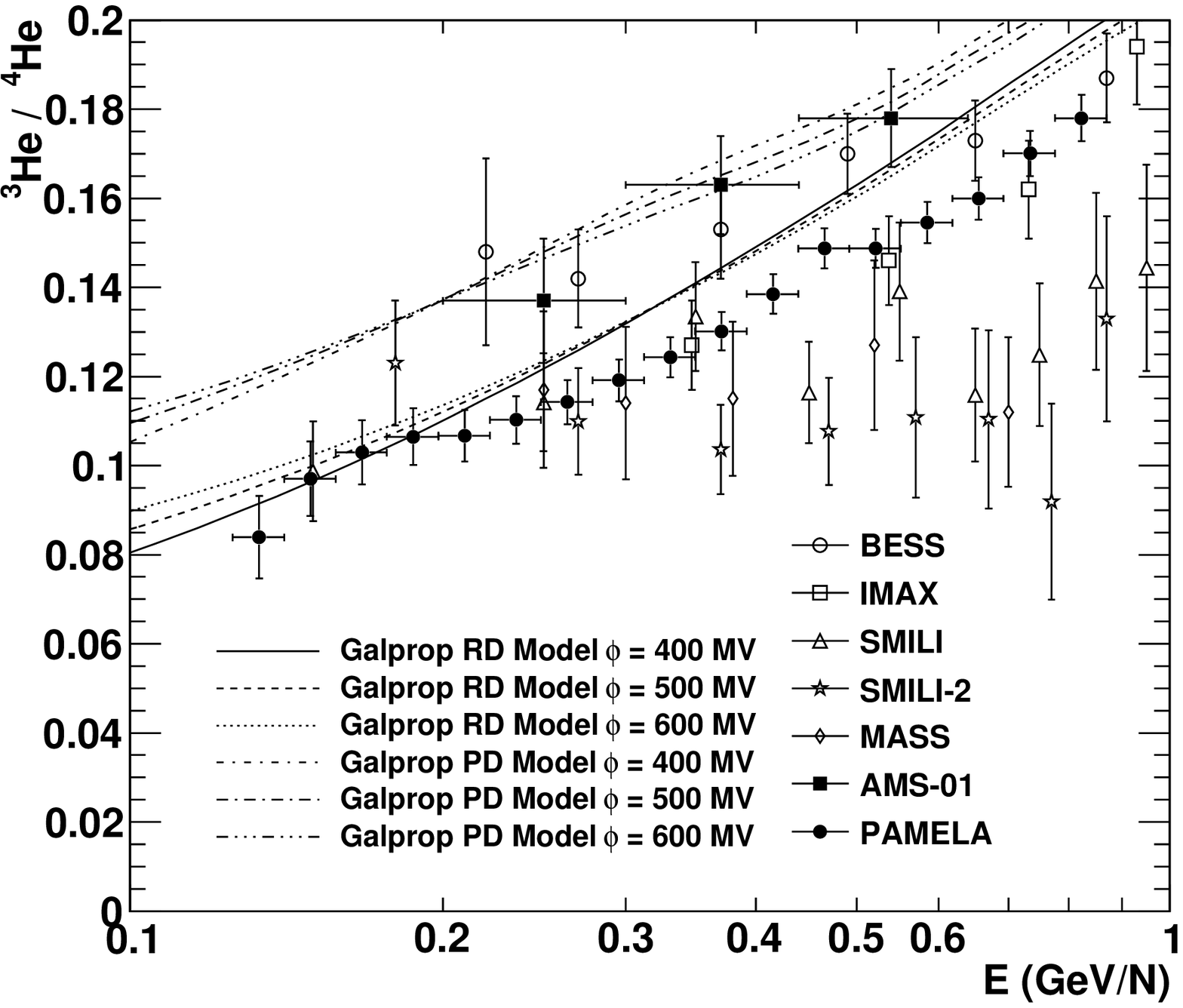}
\hspace{0.5cm}
\includegraphics[width=7.5cm,height=7.5cm]{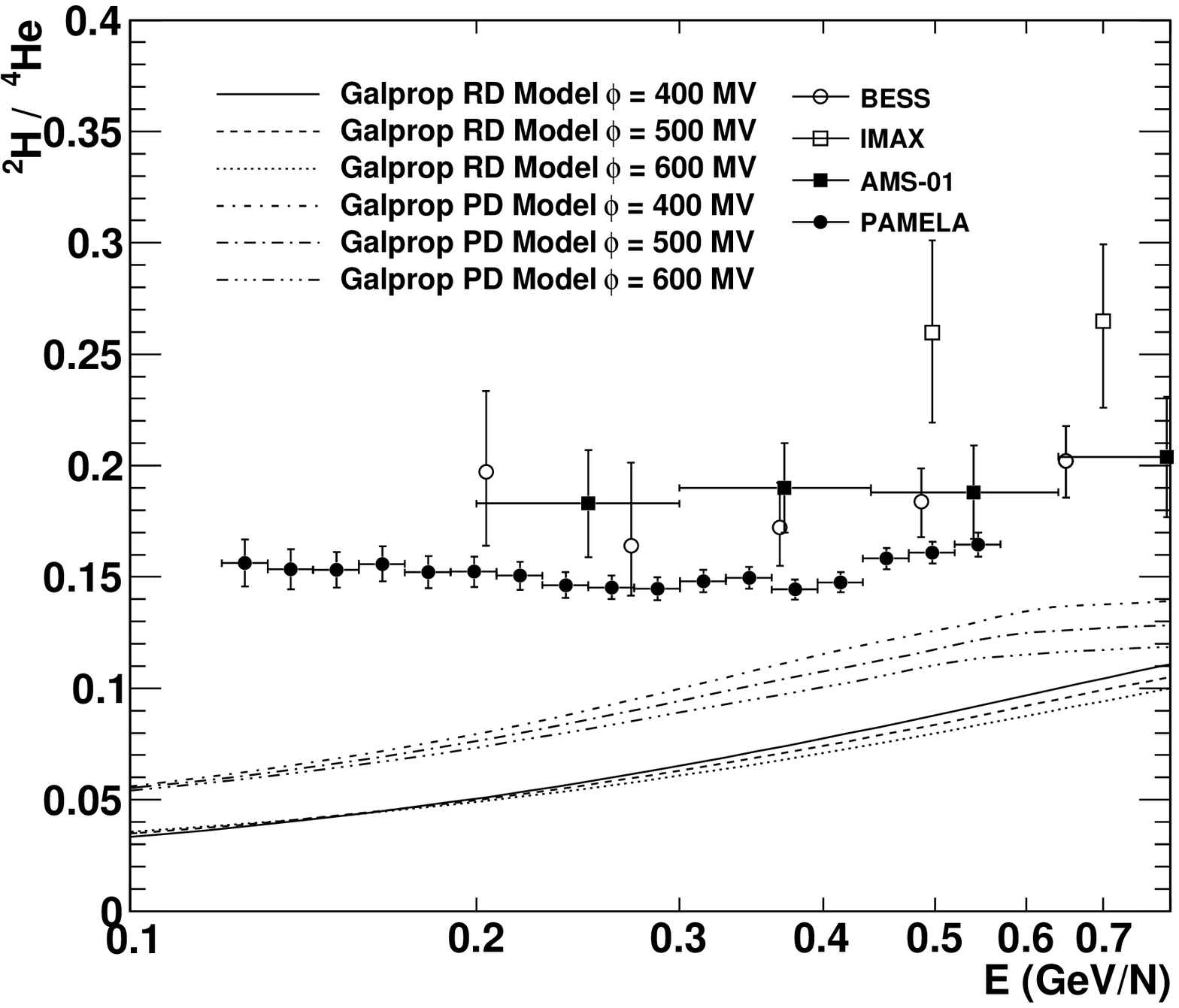} 
\caption{Measurements for the $^{3}He/^{4}He$ (left)
and $^{2}H/^{4}He$ (right) ratios. Most data are from balloon-borne
experiments: 
MASS~\cite{web91}, SMILI~\cite{bea93},
SMILI2~\cite{wef95}, IMAX 92~\cite{den00}, BESS 98~\cite{wan02}
with the
exception of the AMS-01~\cite{agu11} and the preliminary
PAMELA~\cite{for11} results. 
\label{fig:2H3He}}   
\end{figure}

Figure~\ref{fig:2H3He} shows the existing data on the $^{3}He/^{4}He$
and $^{2}H/^{4}He$ ratios.
Isotopes separation was achieved by the mean of rigidity versus time--of--flight selections. As it can be seen most of the measurements were obtained
below a few GeV nucleon$^{-1}$ of kinetic
energy, mostly by balloon-borne
experiments~\cite{web91,bea93,wef95,den00,wan02} and
two~\cite{agu11,for11} 
by space-borne apparatus. In this energy  
domain data are affected by solar modulation and, in the case
of balloon-borne experiments, by a large atmospheric background.

A simple approach for interpreting the data is using the leaky box
model~\cite{web97}. According to this model the propagation of cosmic
rays is described by only one energy-dependent parameter, the escape
mean free path, which represents the mean amount of matter traversed by
cosmic rays before escaping from the confinement volume. Then, the 
comparison between experimental data and the expected abundances of
light secondaries, obtained tuning 
this parameter on the heavier nuclei component,  
is a powerful test 
if their propagation history
is actually the same as that of heavier nuclei. 
Similarly the data can be compared to more sophisticated theoretical
calculations such as those based on the GALPROP code~\cite{str98}
(lines in Fig.~\ref{fig:2H3He} at different solar modulation parameter $\phi$).  
However, the uncertainties on solar modulation 
significantly affects the
interpretation of the results and data would be needed at energies
where these effects are negligible. 
Till now, only one measurement~\cite{pap04} on the
deuterium abundance exist above 10~GeV nucleon$^{-1}$ but still obtained
with a balloon-borne experiment, hence affected by uncertainties on
the atmospheric background. Recently, the AMS experiment~\cite{bat05} was placed on
board the International Space Station. More details will be provided
in Section~\ref{s:future}, but the combination of a magnetic
spectrometer with an acrylic Ring Imaging Cherenkov should extend the
light isotopes measurements up to about 10~GeV/n.

\section{Electron data}
\label{electrondata}

Electrons (and positrons) constitute about 1\% of the total cosmic-ray
flux. While small, this component provides important information 
regarding the origin and
propagation of cosmic rays in the Galaxy that is not accessible from
the study of the cosmic-ray nuclear components due to their differing
energy-loss processes. In fact, because of their low mass, electrons
undergo severe energy losses through synchrotron
radiation in the Galactic magnetic field and inverse Compton scattering
with the ambient
photons. 

Cosmic-ray electrons and positrons are produced as secondaries by the
interactions between cosmic-ray nuclei and the interstellar matter. In
this case, they are the end product of the decay of short-lived
particles (mostly pions via the decay $\pi^{\pm} \rightarrow \mu^{\pm}
\rightarrow {\rm e}^{\pm}$) produced in interactions. Notice that 
the positrons are produced in slight excess. Since the observed positron component 
is of the order of ten percent and less of the electron one 
above a few GeV (e.g. see \cite{adr09b}), the majority
of electrons must be of primary origin. 
As previously stated, X-ray and TeV gamma-ray measurements 
point
unambiguously to SNRs \cite{all97,aha04}
as acceleration sites of cosmic-ray electrons. However, additional
sources of electrons cannot be excluded. Indeed, 
astrophysical objects
such as pulsars, e.g.~\cite{ato95}, or 
more exotic sources such
as dark
matter particles, e.g.~\cite{cir08}, were invoked to explain the
positron fraction measured by PAMELA~\cite{adr09b} and 
are expected to contribute to the cosmic
radiation with roughly equal numbers of electrons and positrons.

More than in the case of the cosmic-ray nuclear component, structures
in the shape of the
electron energy spectrum are expected as a contribution of large
energy losses and, possibly, of the new sources \cite{del10,nis80}.

In the recent years this field has gained greatly from the addition of
new experimental results from ground based~\cite{aha08}, 
balloon-borne~\cite{cha08} and satellite-based
experiments. Especially the results from the space-borne
Fermi~\cite{ack10} and PAMELA~\cite{adr11b} experiments have been
particularly significant.

\begin{figure}[ht]
\includegraphics[width=25pc]{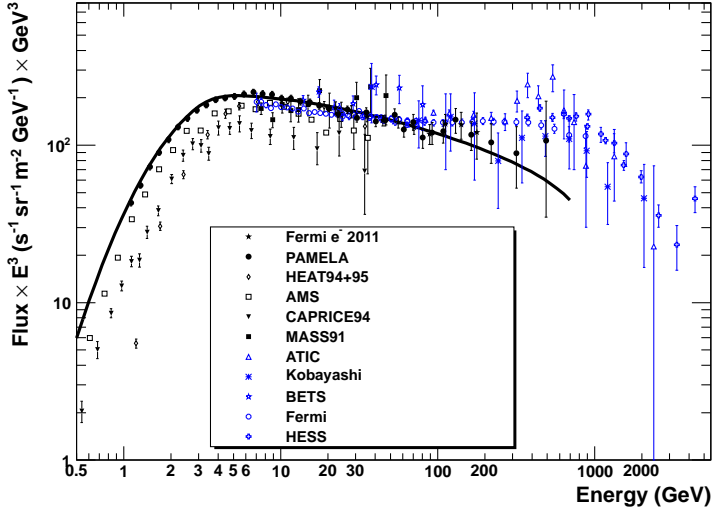}\hspace{2pc}
\caption{The electron energy spectrum from Fermi~\protect\cite{ack10} 
and
  PAMELA~\protect\cite{adr11b}  
with contemporary
measurements:  Kobayashi~\protect\cite{kob99},
CAPRICE94~\protect\cite{boe00}, 
HEAT~\protect\cite{duv01}, BETS~\protect\cite{tor01},
AMS~\protect\cite{alc00}, MASS91~\protect\cite{gri02},
ATIC~\protect\cite{cha08}, HESS~\protect\cite{aha08} 
and a 
theoretical calculation based on the GALPROP code~\protect\cite{str98} 
(solid line). 
Note that the data from
\protect\cite{aha08,cha08,ack10,kob99} and the highest data 
point from HEAT~\protect\cite{duv01} are for the electron and positron
sum.
\label{fig:fluxel1}}   
\end{figure}

Figure~\ref{fig:fluxel1} shows the electron energy spectrum measured 
by PAMELA and Fermi and other recent balloon and space born 
experiments~\cite{cha08,boe00,alc00,duv01,gri02,kob99,tor01}    
and a 
theoretical calculation of the \el spectrum based on the GALPROP
code~\cite{str98}. 
The calculation (solid line) was performed using a spatial Kolmogorov
diffusion with spectral  
index $\delta = 0.34$ and diffusive reacceleration characterized 
by an Alfven speed $v_{A} = 36$~km/s, the halo height
was 4 kpc (parameters
from~\cite{ptu06}). The resulting flux 
was normalized to PAMELA data at $\sim 70$~GeV. For the
secondary \el production during propagation it used primary proton
and helium spectra reproducing the corresponding measured PAMELA
spectra (see Fig.~\ref{fig:PHeflux}) and it was 
calculated for solar minimum,
using the force field approximation \cite{gle68} ($\Phi = 600$~MV).
It should be mentioned that GALPROP does not fully describe cosmic-ray
electron propagation. This calculation is commonly used
assuming a continuous distribution of sources in the Galaxy.
However, due to the
significant energy losses this
does not seem plausible for primary high energy electrons, which  probably originate from a small number of
sources well localized in space.

Both Fermi and PAMELA data show a rather smooth energy dependence of
the energy spectra in a relatively good agreement with the GALPROP
calculation except at higher energies where the experimental spectra
are harder. Such observation was already made by the Fermi
collaboration in their first publication~\cite{abd09}. They concluded
that data and calculation could be reconciled assuming a harder electron
injection spectrum at the source. For example Ahlers et
al.~\cite{ahl09}, following the proposal of Blasi~\cite{bla09b},  
considered the 
production and acceleration of secondaries electrons and positrons by
hadronic interactions of the accelerated protons in SNR shock waves.
With such assumption, they were able to fit the Fermi (and HESS) 
electron data and, at the same time, reproduce 
the increase in the positron fraction measured by
PAMELA~\cite{adr09b}. 
However, additional sources could
also explain the hardening of the electron spectra above 70~GeV and
as well explain the increase in the positron fraction. For more detailed
discussion see~\cite{ser11}.

\begin{figure}[ht]
\includegraphics[width=18pc]{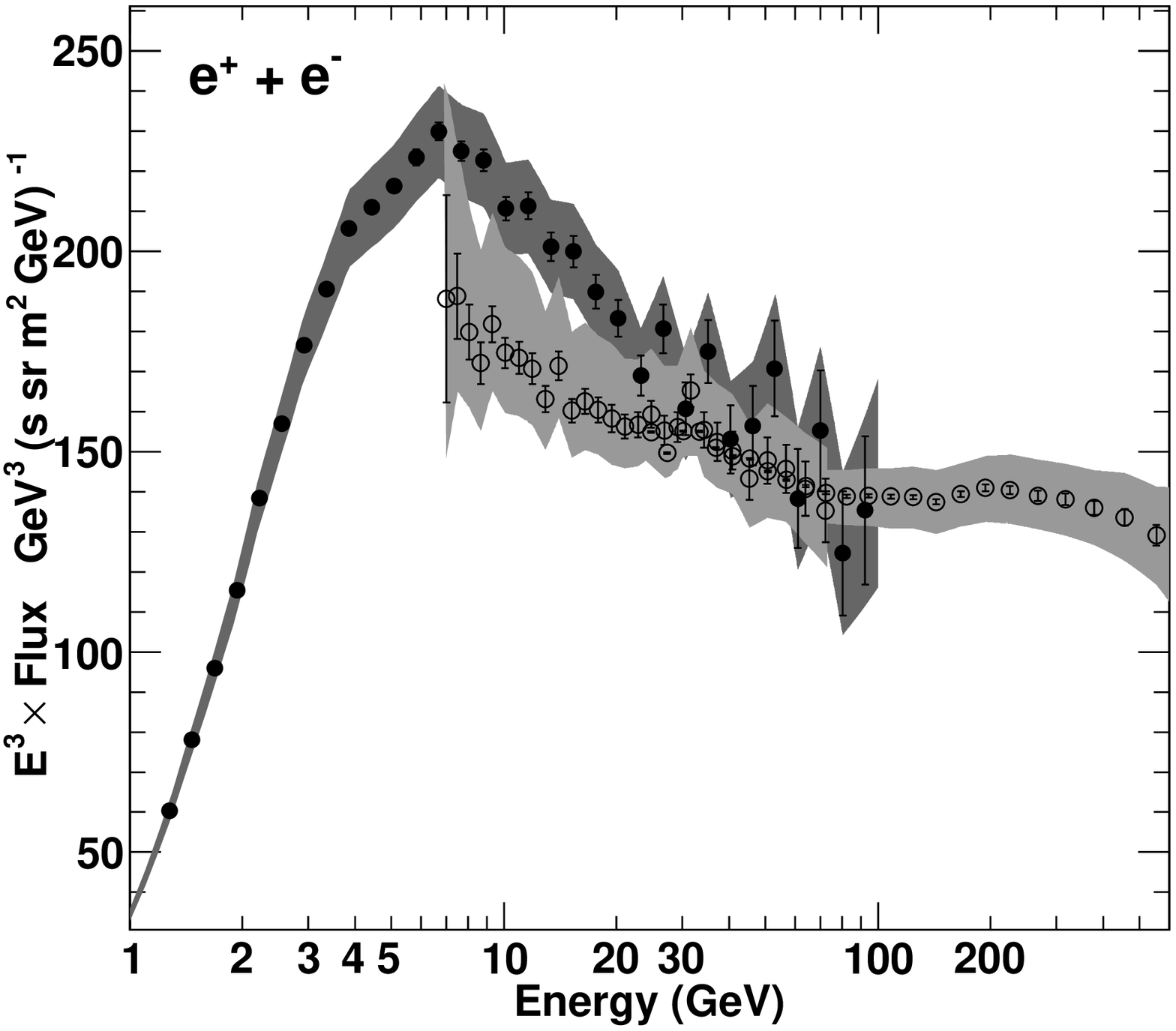}\hspace{2pc}
\includegraphics[width=18pc]{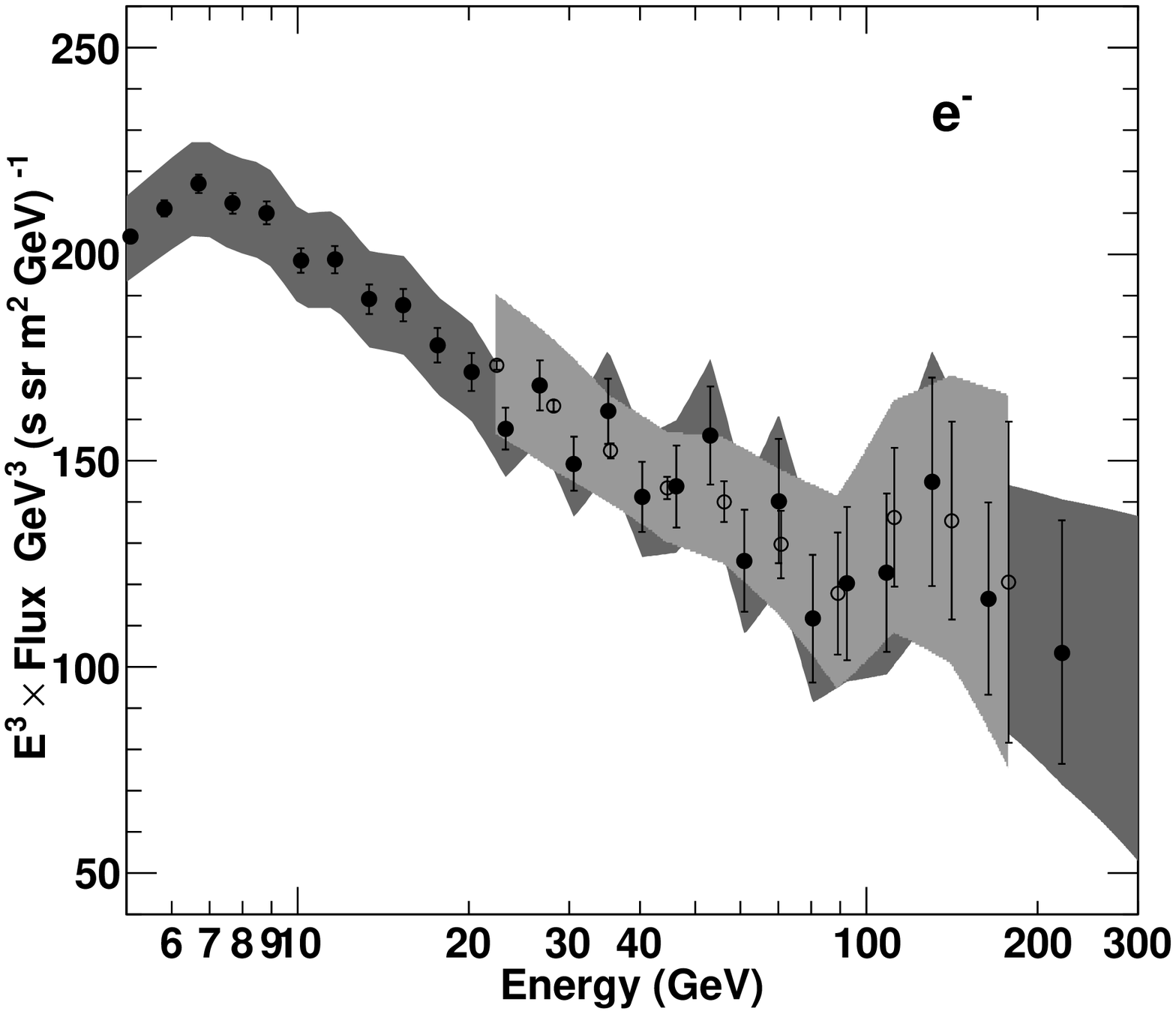}\hspace{2pc}
\caption{Left panel: the all (\el + \ps) electron energy spectrum from
  Fermi~\protect\cite{ack10} (open circles) and 
  PAMELA~\protect\cite{adr11b,adr10a} (full circles). The shaded areas 
represent the overall estimated systematic uncertainties on
the fluxes. 
Right panel: the pure \el energy spectrum from 
PAMELA~\protect\cite{adr11b} (full circles) and 
Fermi~\protect\cite{ack12} 
(open circles), error bars represent the overall estimated systematic 
uncertainties on
the fluxes. 
Because of the presentation of the fluxes multiplied by
E$^{3}$, there are additional systematic uncertainties (not shown here) due to the energy
resolution. These amount to 4\% at 10 GeV increasing to 13\% at 100
GeV for PAMELA and to -10\%,+5\% for Fermi data.}
\label{fig:fluxel2}
\end{figure}

It can be noticed that the Fermi spectrum~\cite{ack10} is
lower as absolute value around 10~GeV than PAMELA data and harder at higher
energies. Therefore, the question of consistency between the two sets
of measurements has to be addressed before a detailed interpretation
of the experimental results. 
First of all, it is important to notice that the Fermi data~\cite{ack10} refer to the sum of electron and
positron fluxes while the PAMELA results refer only to the \el flux. 
Therefore, we constructed a PAMELA ``all electron'' (\el + \ps) spectrum
using the PAMELA \el data~\cite{adr11b} and positron fraction~\cite{adr10a}. Figure~\ref{fig:fluxel2}, left panel, 
shows the Fermi and PAMELA all electron data up to 100~GeV, i.e. the
highest energy bin for which the PAMELA collaboration presented the
positron fraction. The PAMELA spectrum appears 
softer than the Fermi one, however the two spectral indexes differ of
less than a standard deviation ($-3.17 \pm 0.07$ for PAMELA and
$-3.112 \pm 0.002$ for Fermi) when fitting the data with a single
power-law from 30 to 100 GeV and accounting only for statical
errors. The difference in shape is not significant even  
if the two sets of fluxes are increased or decreased in line
with the systematic uncertainties: $-3.17 \pm 0.07$ for PAMELA and
$-3.110 \pm 0.002$ for Fermi and $-3.17 \pm 0.07$ for PAMELA and
$-3.132 \pm 0.002$ for Fermi, respectively. 
At lower energies, the electron flux measured by PAMELA is higher
(about 20\% at 10 GeV) than that measured by Fermi. Considering that
the data were collected partially over the same period of time the
differences cannot be ascribed to solar modulation. 
However, it should be noted that the
systematic uncertainties, shown as shaded areas in
Fig.~\ref{fig:fluxel2}, account for most of the
differences. Therefore, the two measurements can be considered in
agreement. 
Recently the Fermi collaboration released a measurement of the
negative electron spectrum \cite{ack12} showing an improved agreement with
PAMELA results.
Making use of the Earth magnetic field to determine the curvature of cosmic rays (derived from the arrival direction of
the particle into the apparatus, the map of the Earth magnetic field
and the location of the satellite, so called ``East--West effect'') it was possible to determine the sign of the charge of charged particles for 
a subset of Fermi data~\cite{ack12}.
By this method, the Fermi Collaboration was able to measure, 
independently but with lower statistic and in a smaller energy window, 
the $e^-$, \ps fluxes and the positron fraction, see section 
\ref{antidata}. 
Results on the \el spectrum are shown in Fig.~\ref{fig:fluxel2}, right panel, compared to the PAMELA ones. An excellent agreement can be noticed
in the whole overlapping energy range, even considering statistical 
errors only.

\begin{figure}[ht]
\includegraphics[width=25pc]{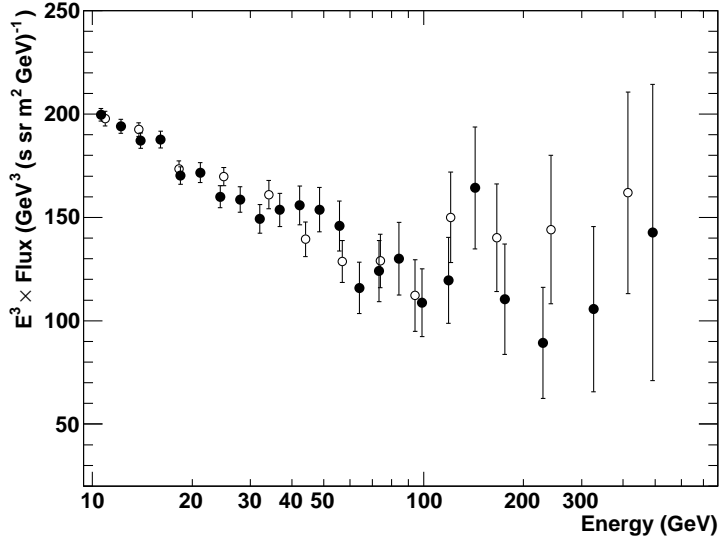}\hspace{2pc}
\caption{The negatively-charged electron spectrum measured by PAMELA
  with two independent approaches: energy derived from
  the calorimeter information (full circles); 
  energy derived from the rigidity (open circles)~\cite{adr11b}. 
  The error bars are statistical only.
\label{fluxtrcal}}   
\end{figure}

As in the case of the hadron energy spectra, differences between the various experimental results might be due
to efficiency and energy determinations. However, differently from the
previous case, as discussed in section~\ref{s:PHe} PAMELA had redundant
determination of the electron (and 
positron) energy. This can be seen in 
Figure~\ref{fluxtrcal} that shows the PAMELA electron 
({\mbox{${\rm e^{-}}$}})
spectra obtained
deriving the energy from 
the calorimeter (full circles)
and the tracking information (open circles). The energy spectrum
derived by the rigidity measured by
the magnetic spectrometer was unfolded to the top of the payload
similarly to what was done with proton and helium spectra, accounting,
however, also for the electron energy losses due to bremsstrahlung while
traversing the pressurized container and parts of the apparatus prior
to the tracking system. In the calorimeter case, tracking information
was used solely to select 
negative particles, thus making a
consistency check possible. Such requirement had no effect on the
energy reconstruction based on the calorimeter data, while losses due
to spillover (i.e. \ps reconstructed as \el and, more significant, \el
reconstructed as \ps) were studied with simulation and found negligible
below 500~GeV.
The two sets of measurements are in good agreement, 
thus validating both the energy determination and the unfolding 
procedures~\cite{adr11b}.

In conclusion, in recent years, especially thanks to new space-borne
experiments, there has been a significant improvement in the
cosmic-ray electron measurements. The recent data, considering all
related uncertainties, are in good agreement both as spectral
shape and absolute value of the fluxes. 
Results, also considering
those expected from
the AMS experiment, are sufficiently
precise to begin for a search of structures expected as contribution
of propagation and single sources (e.g. see~\cite{del10}). This will
be the main scientific goal of future satellite-born experiments such as
CALET~\cite{tor08} and Gamma-400~\cite{gal11} that will have the energy
resolution and acceptance to precisely probe the high energy (100s
GeV-TeV) region.

\section{Antiparticles}
\label{antidata}

Antiparticles (antiprotons and positrons) 
are a natural component of the cosmic radiation 
being produced in the interaction between  
cosmic rays and the interstellar matter. 
Since the first calculations of secondary antiprotons and positrons
(e.g.~\cite{gai74,pro82}) antiparticles 
have been shown to be 
 extremely interesting for understanding 
the propagation mechanisms of cosmic rays. 
 
\begin{figure}[ht]
\includegraphics[width=7.5cm,height=7.5cm]{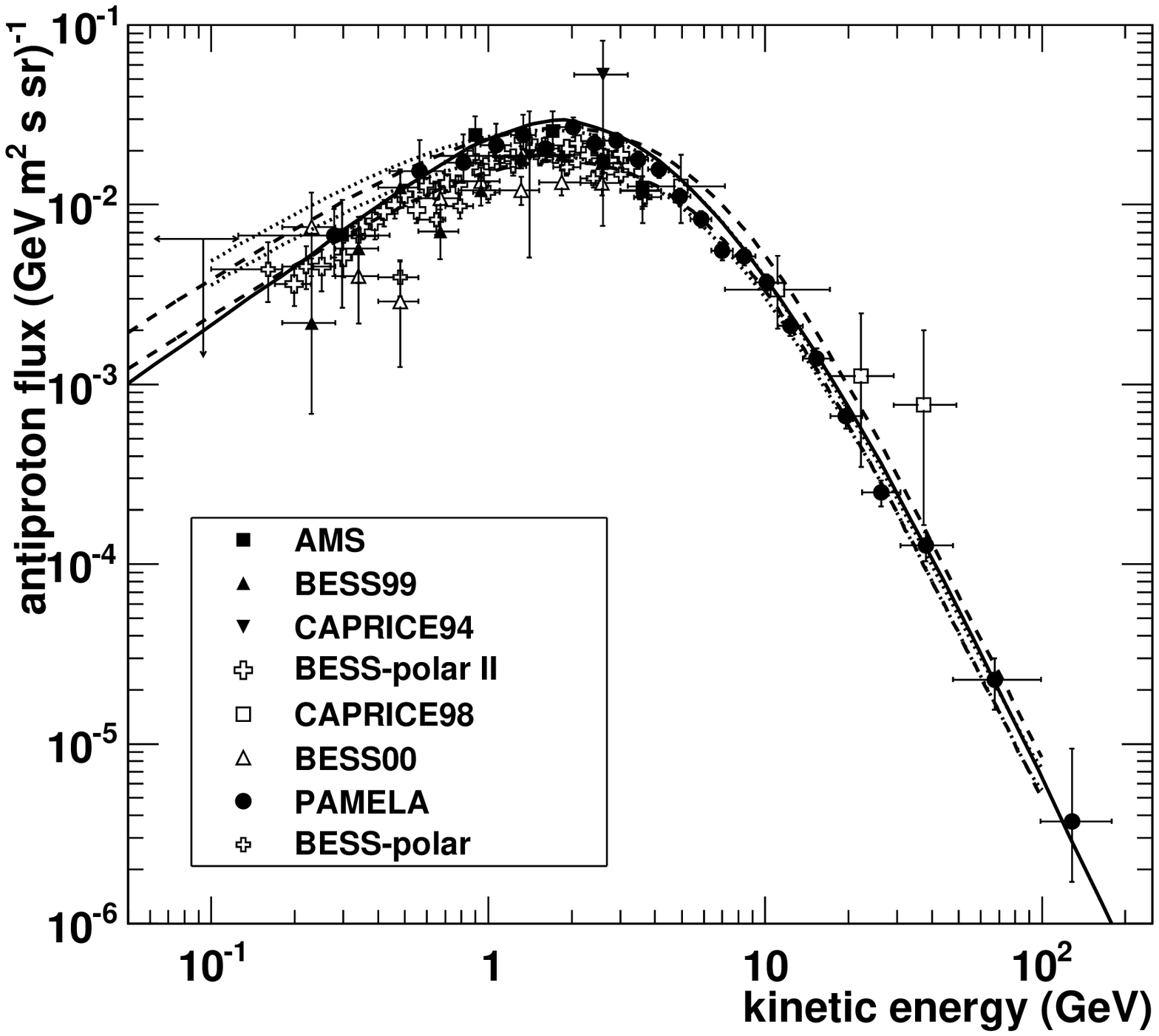}
\hspace{0.5cm}
\includegraphics[width=7.5cm,height=7.5cm]{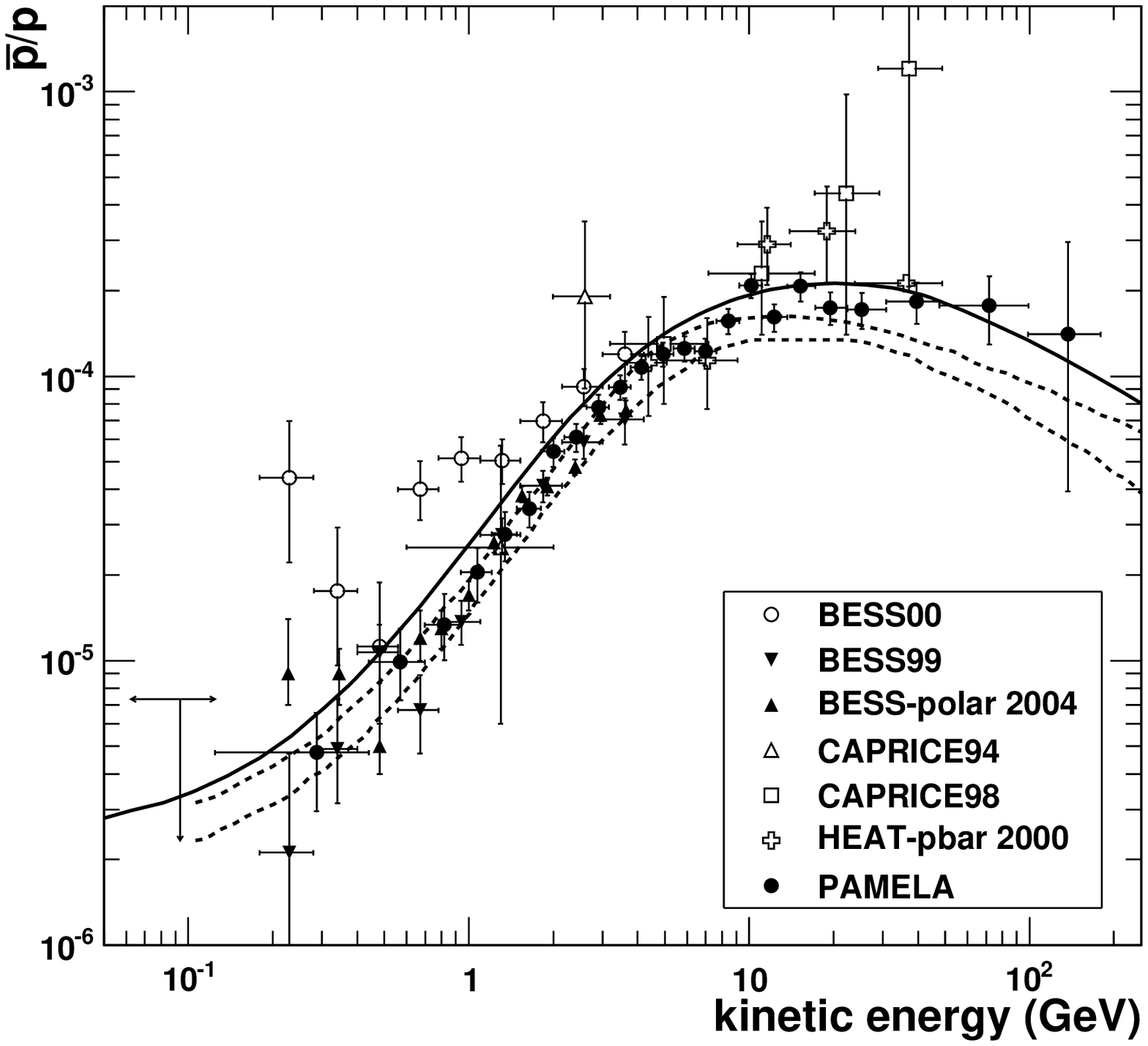} 
\caption{Recent
  measurements of the 
antiproton energy spectrum (left):
PAMELA \cite{adr10b}, CAPRICE94 \cite{boe97}, CAPRICE98 \cite{boe01}, BESS99-00 \cite{asa02}, BESS-polar04 \cite{abe08}, AMS-01 \cite{agu02}, BESS-Polar II \cite{abe12}  and the
  antiproton-to-proton flux ratio (right): 
PAMELA \cite{adr10b}, HEAT-pbar \cite{bea01}, CAPRICE94 \cite{boe97}, CAPRICE98 \cite{boe01}, BESS99-00 \cite{asa02}, BESS-polar04 
\cite{abe08}.
The PAMELA and AMS-01 results are from 
space-borne experiments. 
The dotted lines indicate 
the upper and lower limits calculated by 
\citet{don01,don09} for different diffusion models, including
uncertainties on 
propagation parameters and antiproton 
production cross-sections, respectively. 
The solid line
shows the calculation by \citet{ptu06} for
the case of a Plain Diffusion model. In the multi TeV region
upper limits at $\bar{p} / p \sim 0.1$ from EAS experiments (MACRO~\cite{amb03}, Tibet AS-gamma~\cite{ame07}, ARGO-YBJ~\cite{dis11}) exploiting the moon shadow effect, are not shown in figure.
 The discussion of these results is beyond the scope of this review.
\label{fig:pbar}}   
\end{figure}

Cosmic-ray positrons and 
antiprotons were first observed in pioneering experiments
in the sixties \cite{des64} and 
late seventies \cite{gol79,bog79}, respectively,
using balloon-borne magnetic spectrometers.
While not the only way for detecting antiparticles
(e.g. see~\cite{buf81}) magnetic spectrometers provide a clear and
simple separation between particles and their antipartners.
Figure~\ref{fig:pbar} shows the antiproton energy spectrum (left) and
the antiproton-to-proton flux ratio (right) measured by recent
cosmic-ray experiments 
\cite{adr10b,boe97,boe01,bea01,asa02,abe08,agu02}
along with theoretical calculations assuming pure secondary production 
of antiprotons during the propagation of cosmic rays in the
galaxy.
The experimental data reproduce the falloff below around 2~GeV, 
characteristic of a secondary spectrum, in the
antiproton flux and are in overall 
agreement with pure secondary calculations. However, it may be noted
that the experimental
uncertainties, especially of the recent PAMELA
experiment~\cite{adr10b}, are smaller than the spread in the 
theoretical curves, therefore, pointing to a need for improved 
calculations.

\begin{figure}[ht]
\includegraphics[width=7.5cm,height=7.5cm]{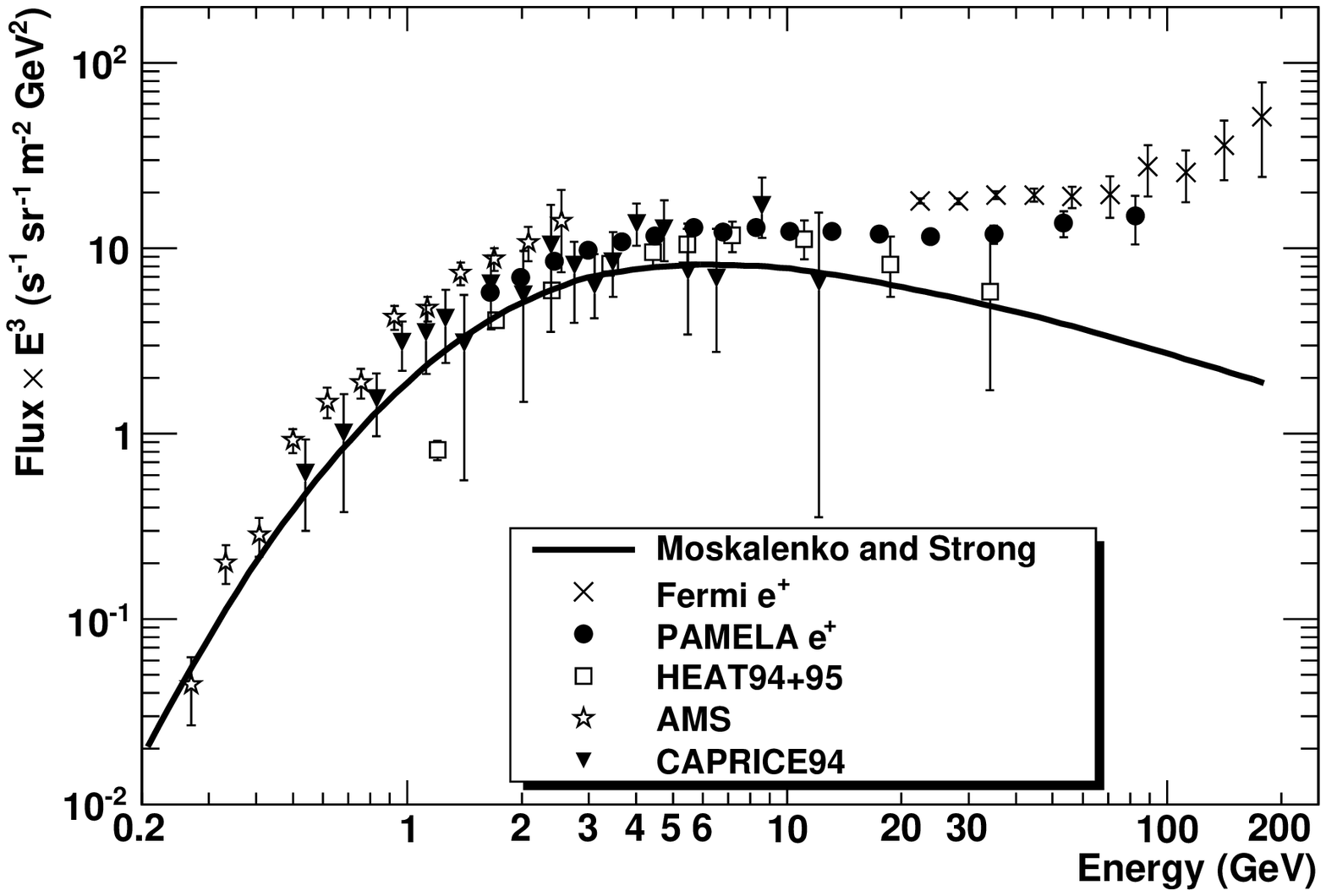}
\hspace{0.5cm}
\includegraphics[width=7.5cm,height=7.5cm]{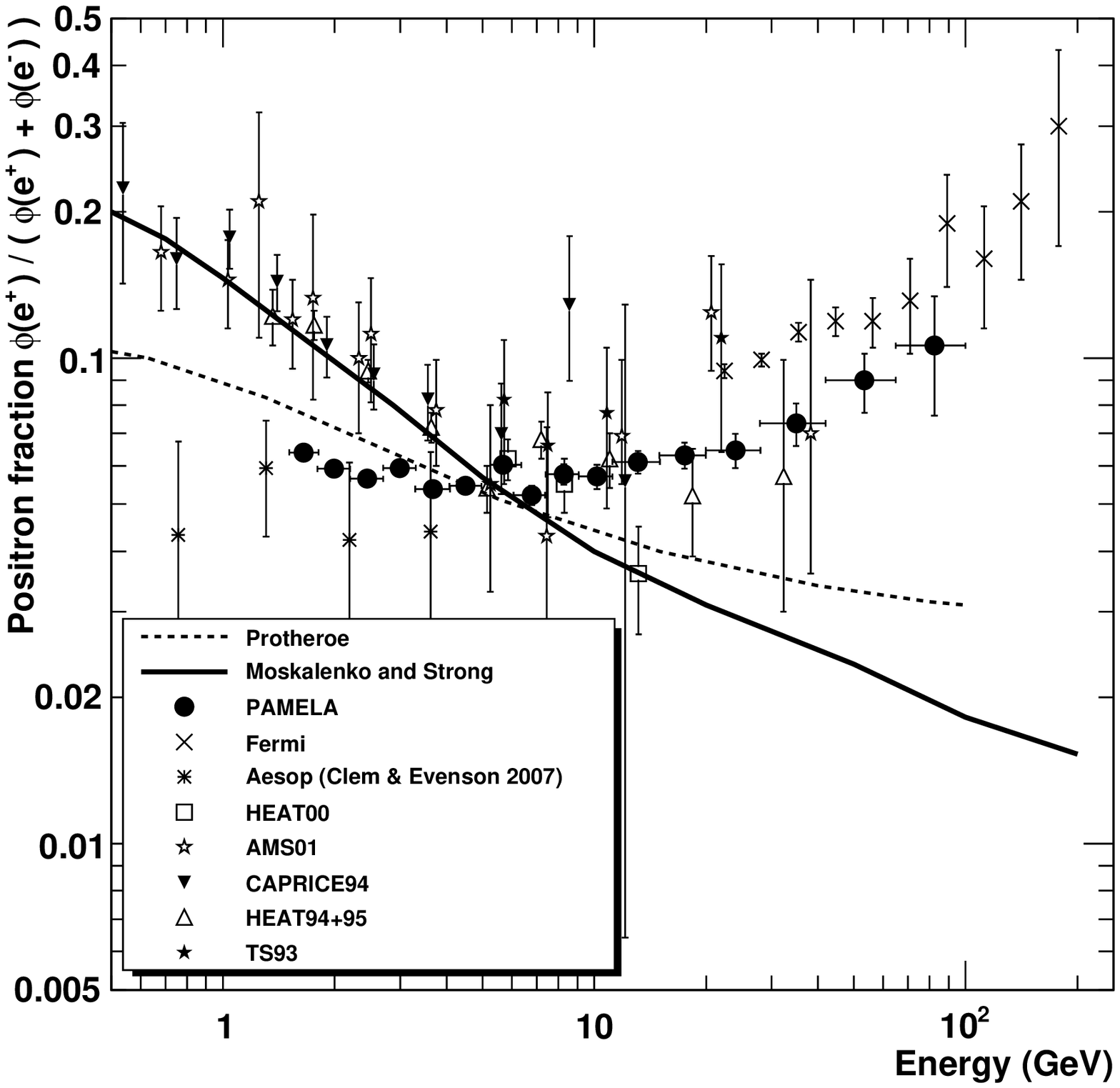} 
\caption{Recent
  measurements
of the 
positron energy spectrum (left):
CAPRICE94~\protect\cite{boe00}, HEAT94+95~\protect\cite{duv01}, 
AMS-01~\protect\cite{alc00}, PAMELA~\protect\cite{adr11b,adr10a}, Fermi~\protect\cite{ack12}
 and the
  positron fraction (right): TS93~\protect\cite{gol96},
  HEAT94+95~\protect\cite{bar97}, CAPRICE94~\protect\cite{boe00},
  AMS-01~\protect\cite{alc00,agui07}, Aesop~\protect\cite{cle07},
PAMELA~\protect\cite{adr10a}, Fermi~\protect\cite{ack12}.
The PAMELA, AMS-01 and Fermi results are from 
space-borne experiments. 
The solid lines show the original GALPROP calculation by Moskalenko \&
Strong \cite{str98} and the dashed line shows the historic calculation
of the positron fraction by Protheroe~\cite{pro82}
for pure secondary production  
of positrons during the propagation of cosmic-rays in the galaxy.
\label{fig:posi}}   
\end{figure}

Figure~\ref{fig:posi} shows the positron energy spectrum (left) and
the positron fraction (right),
ratio of positron and electron fluxes ($\phi$):
\mbox{$\phi$(e$^+$) / ($\phi$(e$^+$) + $\phi$(e$^-$))},
measured by recent
cosmic-ray space-\cite{alc00, adr10a, ack12} and 
balloon-borne \cite{boe00,duv01,gol96,bar97,cle07} experiments.
The solid lines show the original calculation by Moskalenko \&
Strong \cite{str98} (calculated using the force field approximation
\cite{gle68} with solar modulation parameter $\Phi = 600$~MV) and the
dashed line 
shows the historic calculation
of the positron fraction by Protheroe~\cite{pro82}
assuming a pure 
secondary production of positrons during the propagation of
cosmic-rays in the galaxy.
Two features are clearly visible in the data. 
At low energies (below
5~GeV) the contemporary results by PAMELA \cite{adr10a} and by Aesop
\cite{cle07} are systematically  
lower than other data, and at high energies (above 10~GeV) the PAMELA
results show that the positron fraction increases 
significantly with energy opposite to the expectation for secondary
production.  
Below 5 GeV the contemporary measurements of the positron fraction by
PAMELA \cite{adr10a} and Aesop \cite{cle07} are systematically lower than other
data. Above 10 GeV the PAMELA and the subsequent Fermi results show
that the positron fraction increases significantly with energy,
opposite to the expectation for secondary production.
The Fermi results \cite{ack12} 
are in relatively good agreement, considering the systematic 
uncertainties, with PAMELA data.
The measured positron fraction is higher but consistent with PAMELA results and show the same increasing
trend as the energy increases. Partially moderated by the relatively
large uncertainties, this agreement is a positive confirmation of the
results especially for what concern all possible sources of
systematics due to environmental effects.

The low energy discrepancy between the contemporary
results and those from the nineties, i.e. from the previous solar cycle
that favored positively-charged particles, are interpreted as a
consequence of solar modulation effects
(e.g. \cite{lan04,boe09}). The high energy deviation of the
experimental data respect to theoretical calculations may indicate
the contribution of novel sources for cosmic-ray positrons (and
electrons) either of astrophysical (e.g. pulsars \cite{ato95}) or
exotic (e.g. dark matter \cite{cir08}) origin.
 
A detailed discussion of the interpretation of these measurements is
beyond 
the scope of this work and the reader is refereed to~\cite{ser11}.
Instead, here we will discuss the systematic uncertainties related
to the experimental data. 

As for the case of the electron and proton measurements,  
efficiency
and energy estimations are sources of experimental uncertainties,
partially reducible when estimating the ratios or fractions since most
of the efficiencies cancel out. However, their role is less
significant than that played 
by the particle identification in a vast
background of same
charge and sign particles, which are protons for positrons and
electrons for antiprotons for space-borne experiments.
In fact, the antiproton-to-electron and the
positron-to-proton flux ratio 
in the cosmic radiation are approximately 10$^{-2}$ and 10$^{-3}$
between 1 and 100 GeV. Additionally, misidentification of electrons as
positrons and, especially, protons as antiprotons 
can occur if the sign-of-charge is incorrectly assigned from 
the spectrometer data.
 Furthermore, environmental conditions have to be considered. For
example locally produced pions can contaminate the
antiproton sample, especially at low energies. Reentrant albedo
particles may be mistaken for interstellar cosmic rays if the
geomagnetic field of the Earth is not properly treated. 
In the case of 
  balloon-borne measurements, 
  atmospheric muons are a significant background since they are 
  about an order of magnitude larger than 
  the antiparticle signal. Furthermore, atmospheric secondaries,
  i.e. produced by  interaction of cosmic rays with the residual
  atmosphere above the payload, comprise an additional irreducibly
  contamination that can only be estimated by Montecarlo or analytical
  calculations (e.g.~\cite{ste97}) amounting often to 20\% or more of
  the interstellar antiparticle signal in the GeV region.

Several of these backgrounds have different spectral shapes respect to
the signals. Thus, if not completely accounted for, they may mimic
changes in energy spectra and, consequently, produce a
positron fraction or an antiproton-to-proton flux ratio that increases
with energy. 

\begin{figure}[ht]
\begin{center}
\includegraphics[width=25pc]{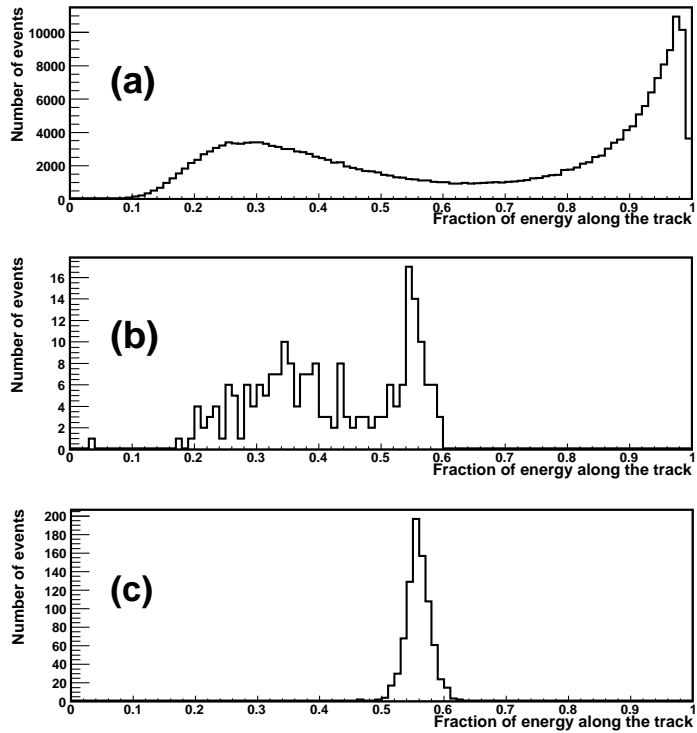}
\caption{The fraction of calorimeter energy deposited
inside a cylinder of radius 0.3 Moli\`{e}re radii in the rigidity
region 28--42 GV. Panel (a) shows the distribution for
positively-charged particles. Panel (b) and (c) show the same
distribution for positively and negatively, respectively, charged
particles selected requiring a match between the momentum
measured by the tracking system and the total detected energy and the
starting point 
of the shower in the calorimeter. See~\cite{boe09} for more details.
\label{fig:enfra}}
\end{center}
\end{figure}

As an example, 
to understand the level of contamination that might lead to a
misinterpretation of the positron data, we analyse one  
energy interval from PAMELA data.
Figure~\ref{fig:enfra} shows 
the fraction of calorimeter energy deposited
inside a cylinder of radius 0.3 Moli\`{e}re radii~\cite{pdg10}
, with axis defined by
extrapolating the particle track reconstructed in the 
spectrometer.  Three cases are shown (see~\cite{boe09} for more details):
\begin{description}
  \item{(a)} positively-charged particles;
  \item{(b)} positively-charged particles selected requiring a match
    between the momentum 
measured by the tracking system and the total detected energy and the
starting point 
of the shower in the calorimeter;
  \item{(c)} negatively-charged particles selected as in (b).
\end{description}
for events with rigidity measured between 28 and 42 GV. 
In this rigidity range and for an acquisition time of about three and half years, with the procedure described
in~\citet{adr10a},  
64.2 positron events were estimated. The expectation according to
GALPROP prediction was for about 18 events. Considering that there
were $\sim 250000$ positive events in this interval (panel (a)), a
residual  
proton contamination of $\sim 2 \times 10^{-4}$ could account
for the larger estimated value. While this contamination value does
not look 
unreasonable, it is worth pointing out that with simple conditions on
the energy deposit in the calorimeter\footnote{Most of the protons do not
interact or interact deeply in the calorimeter, hence depositing low
amount of energy compared to electrons/positrons of comparable rigidity.}
 only 215
positive, panel (b), and 821 negative, panel (c), events remain. As
can be seen in  
Fig.~\ref{fig:enfra}, there is a clear similarity between the
distributions in panels (b) and (c). There are corresponding peaks that can
be naturally associated to positrons and electrons, respectively, and
a broader distribution to lower values in panel (b) that is clearly
ascribed to the residual proton contamination. Assuming that most of
the estimated PAMELA positron signal is due to misidentified protons
means that about two thirds of the associated positron distribution of
panel (b) is due to protons. Moreover, test beam data, simulations and
prior balloon-borne experiments with a similar
calorimeter~(see Supplementary Information of \cite{adr09b}) indicate that 
interacting protons show a rather smooth distribution with no
significant structure for this
variable, therefore making difficult an interpretation of the
distribution at high values of panel (b) as mostly due to
protons. Moreover, the imaging calorimeter provide further information
about the topology of the energy distribution that can be used for
additional rejection. 
It is reasonable to
think that such information can provide an additional factor 100 or
so, thus reducing the residual proton contamination to a small amount
compared to the signal.

All these uncertainties can be studied and partially resolved only by
the use of redundant devices and, especially, multiple measurements
performed by systematically different apparata.
The agreement between PAMELA and Fermi data gives a good confidence that
the increase of the positron flux is indeed to be ascribed to a physical
effect and not to systematic effects affecting the measurements. On the other hand, it could
be argued that an irreducible contamination in the calorimetry
approach for positron identification might affect similarly both the
PAMELA and Fermi results. This issue will be resolved by the AMS
experiment~\cite{bat05} on board the Interstellar Space Station that will employ a Transition Radiation
Detector along with an imaging calorimeter for particle identification.

\section{Nuclei data}
\label{section:nuclei}
The relative abundances of elements and of isotopes heavier than Helium is another essential piece of information to understand 
the origin and history of cosmic rays even if heavier nuclei only
account for about few percent of the total flux of cosmic rays.

A remarkable resemblance between the the galactic cosmic ray source composition and the 
abundances found in the solar system can be notice.  Both cosmic rays and the solar corona/solar wind show evidence of
having undergone chemical alteration, and the resulting fractionation patterns are strikingly similar \cite{stone98}. 

However, it has long been recognized that in cosmic rays the observed abundance
ratio (Li+Be+B)/(C+N+O) exceeds the value found in solar system material
by a factor of about 10$^5$. This difference has been considered to be a measure of how much material
cosmic rays have traversed since they were accelerated. Carbon, Nitrogen and Oxigen are considered primary cosmic rays, i.e. nuclei that
are produced and accelerated by sources and reach Earth without undergoing fragmentation, while Litium, Berilium and Boron are secondary components resulting from 
fragmentation reactions in the interstellar medium.
Galactic cosmic ray sources and propagation models can be studied 
measuring the primary and secondary nuclei in the cosmic rays.

In addition to stable isotopes, the cosmic rays contain long--lived radioactive nuclides of either primary or secondary origin.
The observed abundances of these ``clock'' isotopes can be used for establishing various time scales related to the origin of cosmic
rays:
\begin{itemize}
\item primary isotopes can shed light on the nucleosynthesis process in the cosmic-ray sources.
Primary isotopes measured in the cosmic rays have undergone some type of chemical fractionation. It
is of major interest studying these isotopes to understand how and where this has occurred \cite{mewaldt01}; 
\item primary isotopes which decay by electron
capture can provide information about the elapsed time between nucleosynthesis and particle acceleration. These types of isotopes
are usually called ``acceleration delay clocks''\cite{mewaldt01}; 
\item secondary isotopes which decay by $\beta^{\pm}$ emission
can measure the average time between the production of these particles
and their escape from the Galaxy (``propagation clocks'') \cite{ptuskin98};  
\item secondary isotopes which instead decay solely by electron capture can do so only if their velocities have, at some time, been
much lower than the velocities at which they are observed. For this reason they
are potentially probes of energy changing processes that occur during
propagation in the Galaxy (``reacceleration clocks'')\cite{mewaldt01}.
\end{itemize}

Hence high resolution measurements by cosmic ray nuclei detectors provide very important
clues to the astrophysics of our Galaxy and can
narrow the range of possible explanations for the origin and propagation history of galactic cosmic rays.

\begin{figure}[ht] 
\includegraphics[width=25pc]{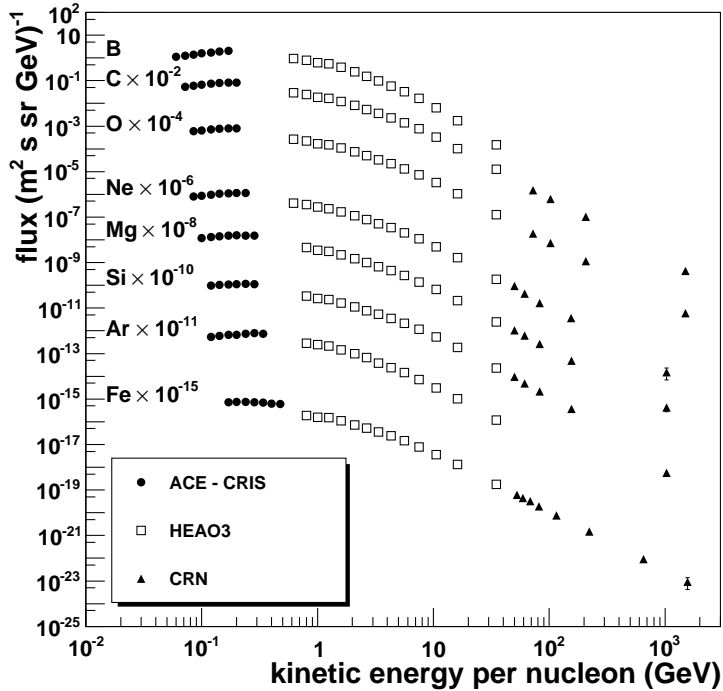} 
\caption{Observed spectra for a selected set of elements as measured by space experiments CRN \cite{muller91}, HEAO--3 \cite{engelmann90}, ACE (at solar minimum) 
\cite{george09}. To reduce overlap the spectra for individual elements are shifted in vertical direction as indicated.
\label{fig:nuclei}} 
\end{figure} 
Figure \ref{fig:nuclei} shows recent nuclei spectra measurements made by the space--based detectors CRN \cite{muller91}, HEAO--3 \cite{engelmann90}, ACE (at solar 
minimum) \cite{george09}. Notice that the study of a wide energy range nuclei spectra and elemental composition has been carried out mainly by balloon--borne experiments (for example CREAM \cite{ahn08}, ATIC \cite{pan07} and TRACER \cite{ave09}, not shown in figure). When considering the whole existing data set, a good agreement can be seen between different measurements. At the highest energies the measurement is usually limited by the statistics. 

\begin{figure}[ht]
\includegraphics[width=15pc]{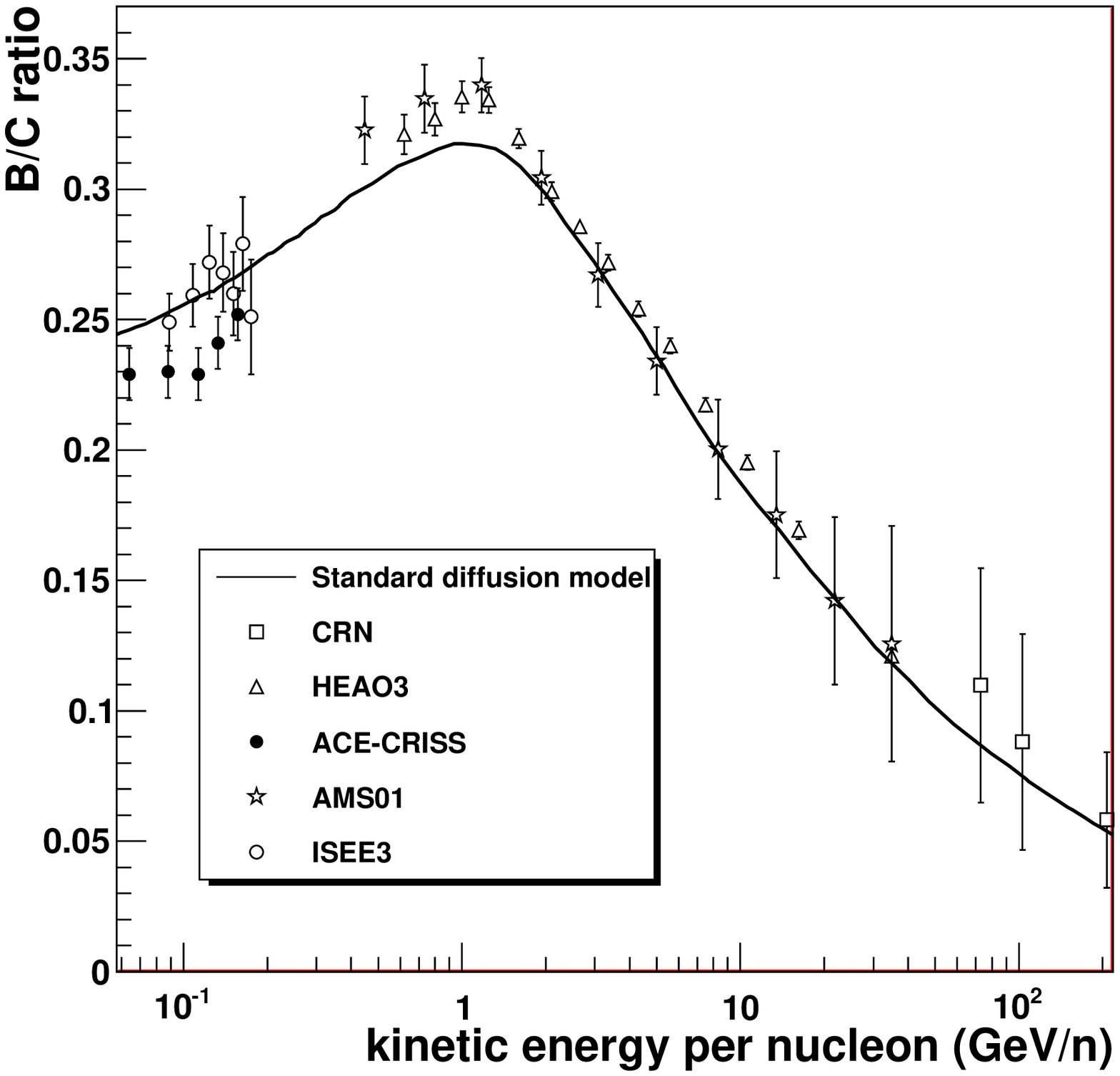}
\includegraphics[width=15pc]{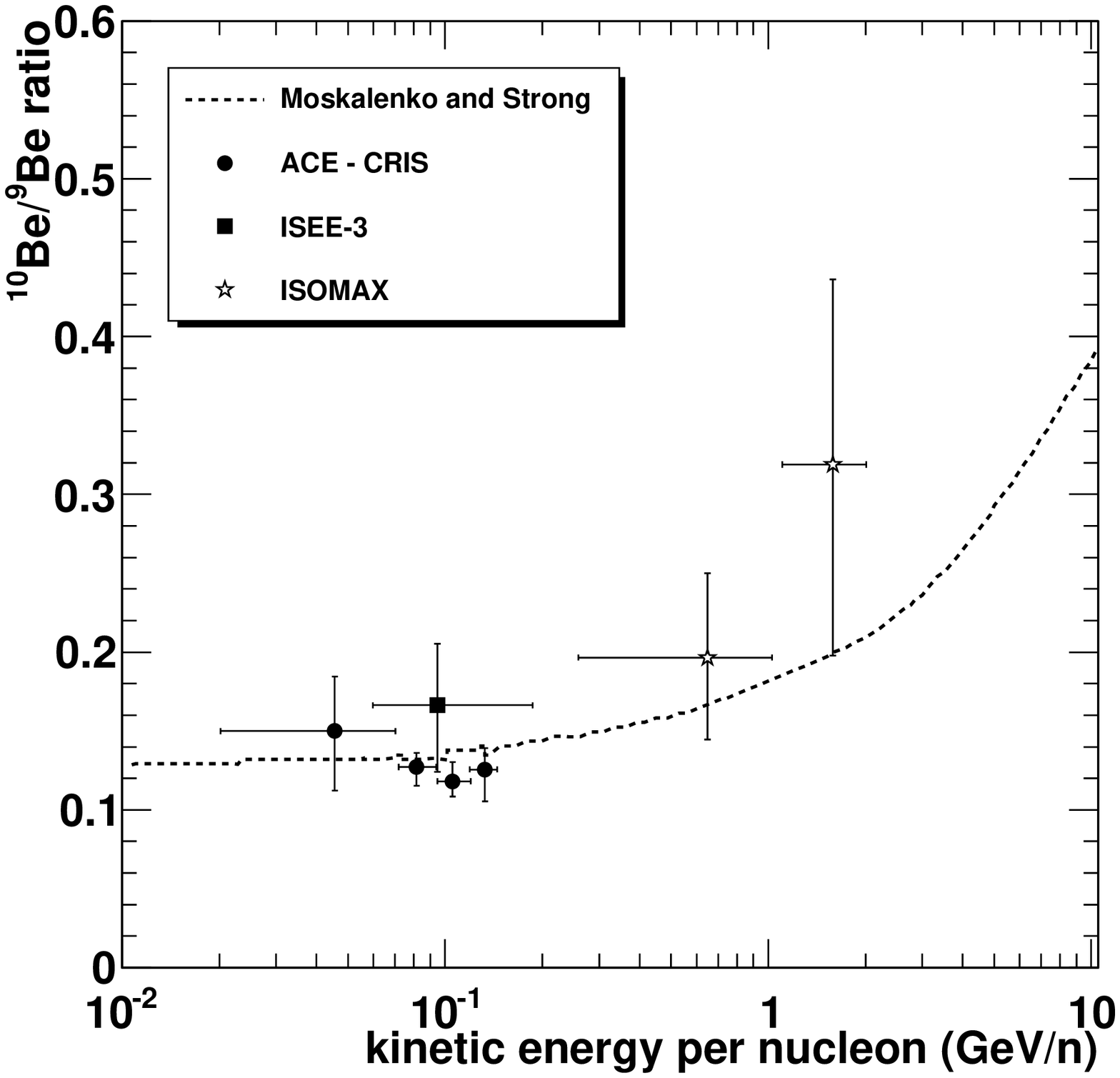}
\caption{On the left, the B/C fraction as measured by the space experiments ISEE--3 \cite{les93}, CRN \cite{muller91}, HEAO--3 \cite{engelmann90}, ACE \cite{davis00}, AMS01 \cite{tom11}; solid black line is a fit to the data assuming a standard diffusion model \cite{strong07}. 
On the right the $^{10}$Be/$^{9}$Be ratio as measured by space experiments ACE \cite{davis00} and ISEE--3 \cite{wie80} and the balloon--borne experiment ISOMAX \cite{ham04}; 
the dotted black line is a diffusive halo model with 4--kpc scale height using GALPROP \cite{strong01}. \label{fig:bc}}   
\end{figure}
In Fig. \ref{fig:bc}, left panel, the B/C ratio is shown. Since the Boron is expected to be a pure secondary component
this ratio is considered a reference for this type of measurements. Moreover, since the major progenitors
of B are C, N, and O the production cross sections are better known respect to other secondary nuclei. 
Solid line in figure represents a fit to the data assuming a Galaxy leaky box model where the chemical
composition at sources resembles the solar like abundances. The set of parameters found fitting B/C is in general consistent
with all other secondary/primary ratios. Different models (standard or turbulent diffusion models, wind model, minimal re-acceleration model) cannot be distinguished 
by these data alone. All these models can be used to determine the injection spectral index of primary nuclei which
is found to be in the range 2.3 -- 2.4 for C and Fe in the energy range 0.5 -- 100 TeV \cite{strong07}.

The break at low energies in the B/C ratio is a puzzling feature for which many explanations have been proposed. It has been claimed that
HEAO--3 data combined to Voyager 2 show that the break is due to a solar modulation effect \cite{webber03}.
Another possibility is the reacceleration of nuclei which can explain the shape of the B/C ratio with a plausible fit and without the
need of ad--hoc break in the diffusion coefficient. The reacceleration models are the favoured ones, since 
some level of reacceleration is expected when diffusion occurs on moving scatterers. Another simpler explanation of
the B/C energy dependence is the local--source model \cite{davis00, moskalenko03}, in which  a local component 
of primary cosmic rays is assumed. Since the contribution of local secondaries can be considered negligible, a steep local primary source
can eventually make the B/C decrease at low energy. 

Figure \ref{fig:bc}, right panel, shows the $^{10}$Be/$^{9}$Be ratio compared to the diffusive halo model obtained using GALPROP. 
This type of radioactive secondaries (propagation clocks) can only 
travel a few hundred parsec before decaying; by the mean of these measurements it is possible to determine the diffusion coefficient.
Assuming that the diffusion coefficient does not vary from the local region to the full galaxy volume,
using the secondary to primary ratio, it is possible to estimate the size of the full propagation region.
Given the present measurements, typical results are $D_{xx} \sim 4 \times 10^{28}$ cm$^2$ s$^{-1}$ (at 3 GV) for the diffusion coefficient and z$_h$ = 4 kpc \cite{strong07} for the height of the propagation region.

A comprehensive set of ``acceleration delay'' and ``reacceleration'' clocks have been measured by the ACE detector. Results on the Co and Ni isotopes seem to be
consistent with a delay $\ge 10^5$ y from the synthesis to the acceleration. This observation is not in agreement with models
in which the supernova accelerate their own ejecta and seems pointing to the acceleration of existing interstellar matter \cite{strong07}.
Results on reacceleration clock isotopes, like V, Cr and Ti, usually seems to be better in agreement with models including reacceleration, however
measurements are not always consistent \cite{strong07}. 

Nuclei measurements have been performed over the years using different techniques: energy loss detectors (ISEE--3 \cite{greiner78}, ACE-CRISS \cite{stone98}), Cherenkov detectors (HEAO--3 \cite{bouffard82}), Transition Radiation Detector (CNR \cite{swordy90}), magnetic spectrometers (AMS-01 \cite{alc00}, PAMELA \cite{pic07}).  

The nuclei charge identification is usually achieved by measuring the ionization losses 
in scintillators or silicon detectors, but also other tecniques have been 
adopted, for example Cherenkov counters or TRDs. Experiments dedicated to the study
of nuclei have redundant charge detectors that make the nuclei selection very clean.
However, more challenging is the efficiency determination of the particles selection. 
Nuclei, in fact, can interact within the detector not only producing a hadronic shower 
but also undergoing a fragmentation into lighter nuclei. The ability and the
efficiency of the detector in discarding these type of events can usually only be 
tested making use of simulations or test beam data. Systematic uncertainties are hence introduced,
since simulations suffer the sometimes poor knowledge of cross--sections for 
``uncommon'' nuclei in the desired energy range, while test beam data must be tuned
to nominal working conditions of flight detectors and again it is not always possible
to have beams of the full range of nuclei and energies.

The nuclei energy measurement technique depends on the energy range that has to be covered. At low energy, up to about 1 GeV/n,
non--interacting stopping nuclei are selected and their energy released measured making use of homogeneous calorimetry measurements.
The main issues of this approach are the ability of the detector of selecting pure samples of non--interacting particles and the efficiency
determination of this selection. Nuclei spallation can potentially be a source of contamination of lower Z nuclei from higher Z ones when the 
nuclei selection is based only on the energy release. For example it can be difficult to distinguish a Boron nucleus
from a Carbon which by spallation has lost a proton which is traveling along the same trajectory -- Carbon has charge six, 
 it would release an energy proportional to $Z_C^2$ but losing a proton it will become a Boron releasing energy proportional to $Z_B^2=(Z_C-1)^2$ plus a proton releasing order of one MIP per detector for a total of $(Z_C-1)^2 + 1$ which has to be compared to the release of $Z_B^2=(Z_C-1)^2$ of a single primary Boron.

In the same energy range also the time of flight has been used to determine the nucleus velocity. Slow--down effects in this case must be
taken into account and are a source of systematic uncertainties.
At higher energies, from the GeV/n to tens of GeV/n, the energy can be measured using Cherenkov detectors and  the logarithmic rise of
ionization losses detectors, the latter one providing a low energy resolution. A Cherenkov detectors hodoscope has been used in the HEAO--3 experiment \cite{engelmann90}. Providing
multiple energy measurements an excellent control of systematics errors has been obtained.
At the higher energies transition radiation detectors and magnetic spectrometers have been used. 
In the case of magnetic spectrometers the nuclei energy measurement requires a dedicated calibration of the detector. Usually the high level 
emission of delta rays while passing through position measurement detectors can induce a distortion of the recorded cluster which must
be identified and taken into account. Moreover, high energy releases, for high Z or very slow nuclei, can saturate the detected signal worsening 
the position resolution and hence the energy resolution.
At the highest energies deep calorimeters must be used. At very high energies nuclei interaction cross section become not--negligible and
most of the nuclei interact when entering a calorimeter. Energy released is usually proportional to the incoming particle energy for very deep homogeneous
calorimeters. Weight and space constraint of space experiment usually force the use of calorimeters of limited dimension. In this case
transversal and longitudinal leakage must be properly taken into account.
Indeed multiple energy measurements are needed for dedicated nuclei experiments in order to cover the largest possible energy window and
being able of performing an energy cross calibration of detectors with different systematics.

The space-borne experiment PAMELA was also designed to study 
the abundances and
composition of light cosmic rays (up to oxygen) over almost three decades of energy.
The PAMELA time of flight (ToF) consist of three scintillator double layers
which enable independent charge determinations. Multiple energy deposit measurements combined
with the velocity determined by the ToF and with the nuclei rigidity as 
measured by the tracking system can provide redundant information that
improve significantly the charge resolution.

Moreover, even if the PAMELA instruments is optimized for the
detection of positrons and antiprotons, three different detectors (ToF, tracker and
calorimeter) are able to identify, with different efficiencies, resolutions and Z ranges, light nuclei. 
Hence it is possible to perform a highly accurate charge measurement by selecting particles
independently with the three detectors. The nuclei spectra can be measured up to different momentum depending on the 
charge since the tracking system resolution depends on the nuclei rigidity.

\begin{figure}[!ht]
\includegraphics[width=32pc]{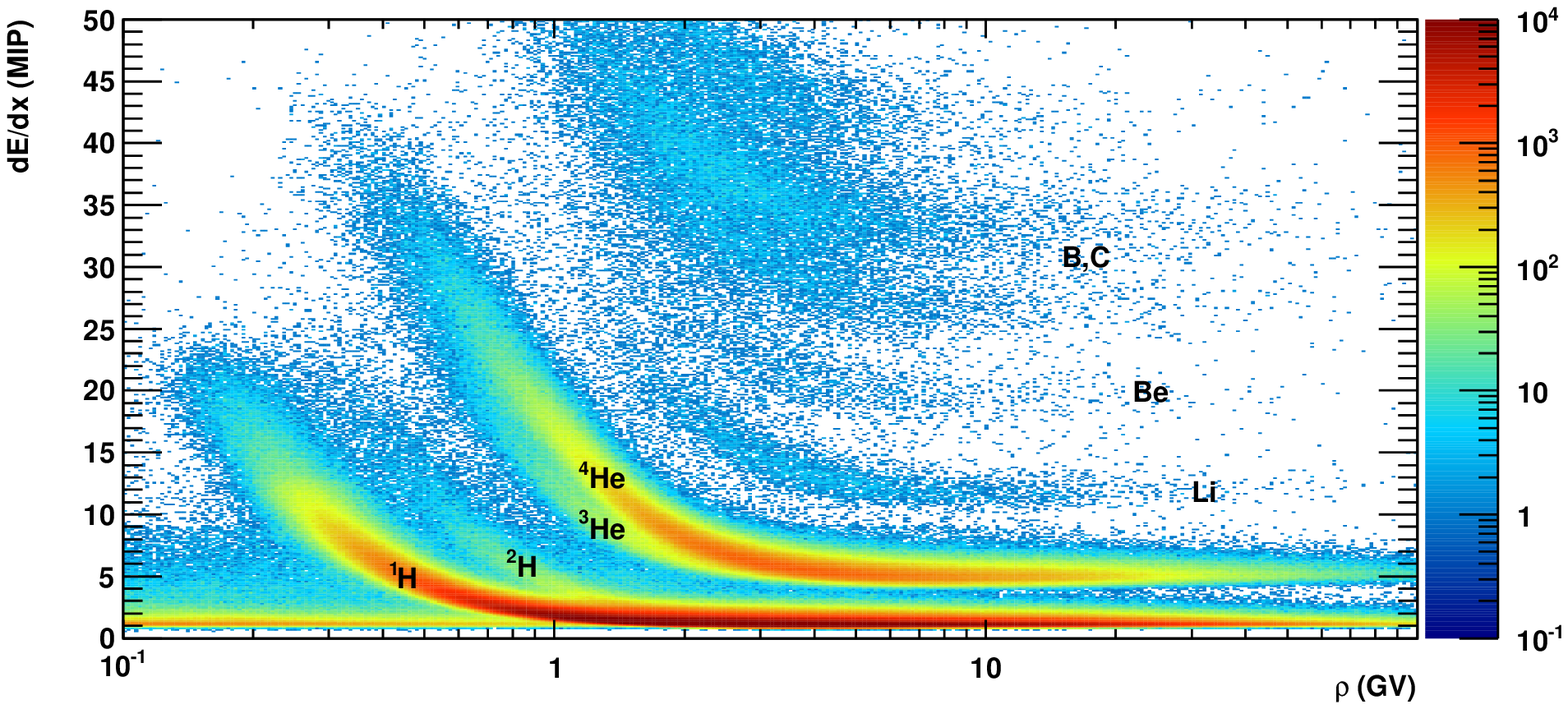}
\includegraphics[width=32pc]{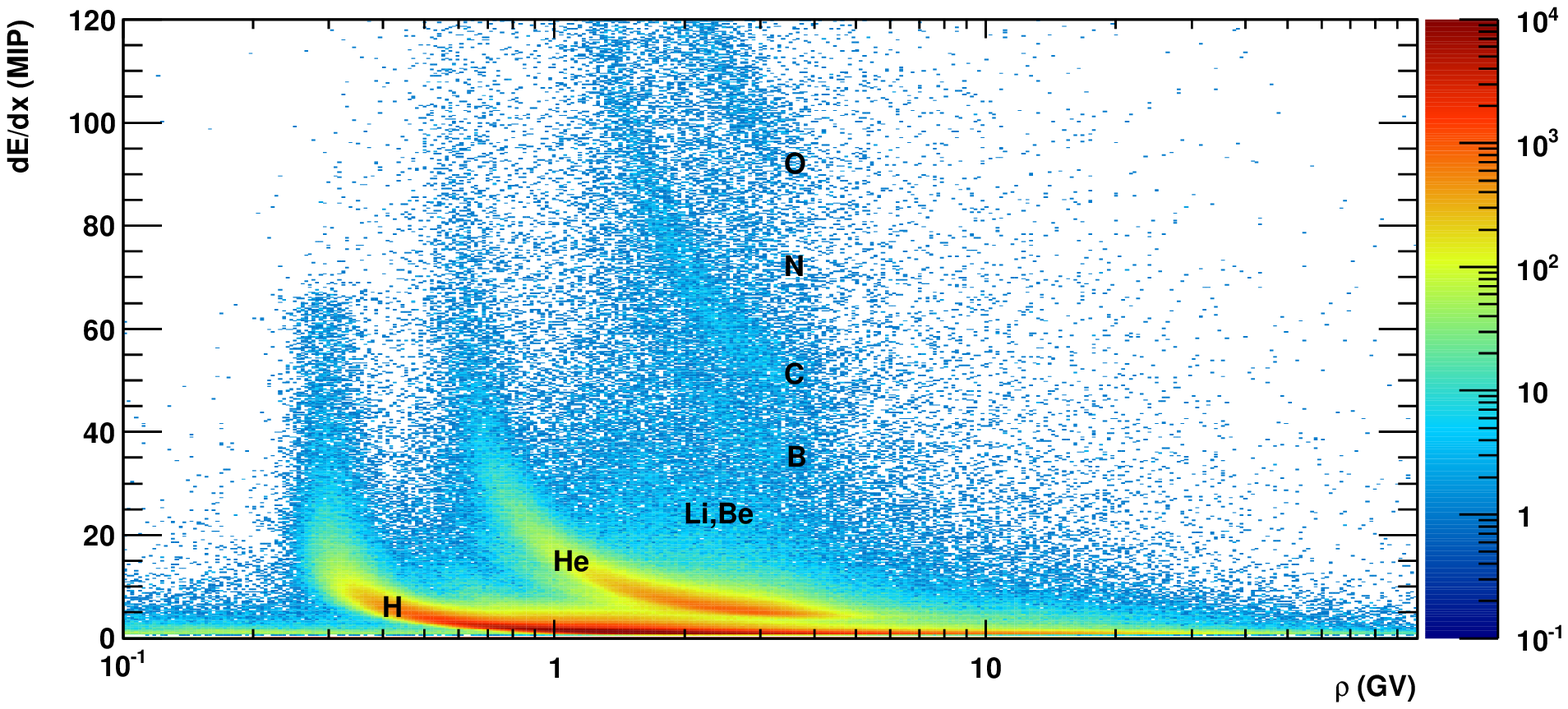}
\caption{The nuclei separation capabilities of PAMELA ToF (top) and calorimeter (bottom) \cite{for11}. }
\label{fig:nucleibands}   
\end{figure}
By comparing the ToF charge measurement with the particle rigidity
the different nuclei separated into different bands, as shown in Fig. \ref{fig:nucleibands}, on the top. 
By fitting these bands and projecting the signal along the fit lines it is possible to determine the charge resolution.
The charge gaussian distribution has a standard deviation that is less than 0.1 for protons and 0.16 for C (in units of proton charge e). A good charge separation
can also be obtained comparing the energy released on the first calorimeter plane versus the rigidity as measured by the
tracking system, Fig.  \ref{fig:nucleibands}, on the bottom. In this latter case charges higher than Oxygen can be separated since the dynamic range in MIP of the calorimeter 
strips is much bigger than the ToF and tracker ones.

Below $\sim$2 GeV/n, three independent energy measurements are 
available: the time of flight, the deflection of the particle in the
spectrometer magnetic field, the Bragg's peak of the nucleus stopping in the 
calorimeter. In this energy region is therefore possible to put constraints
on the tracking system systematics errors in the energy measurements. 
Moreover a cross--check of nuclei selection efficiencies as function
of the nuclei energy can be used to estimate systematics in the nuclei separation.
To extend the nuclei measurement at higher energies (hundred of GeV/n)
the tracking system must be used to measure the particle rigidity.

The PAMELA B/C ratio will be a very important measurement to put strong
constraints on cosmic--ray propagation and acceleration models.
In fact, the use of a full data set (matter, anti--matter and nuclei over a wide energy range) provided by a single instrument permit to
avoid inconsistencies between data sets from different experiments 
and minimize uncertainties on the solar modulation parameters which is 
difficult to parametrize properly. 

\section{Future}
\label{s:future}

On the 19th of May 2011 the AMS-02 apparatus~\cite{bat05} was
installed onboard the 
International Space Station (ISS) and it started collecting data. The
apparatus resembles the PAMELA one being equipped with a permanent
magnet, a silicon tracking device and an electromagnetic
calorimeter. However, AMS has a significantly larger acceptance (about
a factor 20) and additional detectors such as a Transition Radiation
Detector and a Ring Imaging Cherenkov detector that will provide
a significant improvement in statistics and systematics respect to
PAMELA concerning antiparticle and chemical composition of the cosmic
radiation. 

Another experiment designed to study the electron component and the
chemical composition of the cosmic radiation with a calorimetry approach
is CALET~\cite{tor08}. The apparatus is built around a 30 radiation length
calorimeter and it 
will be placed on board the ISS sometime around 2014. Major scientific objectives are to search for nearby cosmic ray sources and dark matter. With an
acceptance of about 0.12 m$^{2}$sr, CALET will be able to precisely 
measure the
all-electron energy spectrum from 20 GeV to 10 TeV.
Though not optimized for hadrons, CALET has also a capability to measure protons and nuclei up to 1000 TeV, and will have a function to monitor solar activity and $\gamma$ ray bursts with additional instrument. CALET expects to measure $\sim$ 20 protons and $\sim$ 15 nuclei of the iron family above 5$\times$10$^{14}$ eV in a five years mission.

A similar
calorimetry approach will be employed by Gamma-400~\cite{gal11}
to study the
high-energy gamma-ray flux and cosmic-ray  electrons and nuclei.
The apparatus will be placed on board a Russian satellite, which launch
is foreseen for 2017-2018. With a similarly deep but significantly
larger calorimeter (acceptance of about 1 m$^{2}$sr), Gamma-400 will
be able to increase by about an order of magnitued the statistics acquired by CALET.

The energy region close to the knee (located at $\sim$3-4 PeV) turns out to be very difficult to explore with balloons or space borne instruments, due to the very low cosmic ray fluxes (2-3 particles per m$^2$sr yr for E $>$ 10$^{16}$ eV) and the need to provide an energy measurement with a mass limited instrument. 
To obtain a statistically significant data sample to study the spectral index change, across the knee region, requires a large collection power and it is only possible with a long duration space experiment.

Taking advantage of a long observation time and a quite large geometric factor, Gamma-400 can extend spectral measurements and studies of cosmic ray elemental composition in energy and provide high precision data at lower energies.

All together the new set of future measurements with higher statistics and in a wider energy range will provide an important tool for testing the theoretical scenarios developed to explain the data published recently by PAMELA, Fermi and contemporary balloon-borne experiments. 

A even more important contribution to the understanding of present day cosmic ray measurements will come from observations and discoveries made with man made accelerators, like LHC. If new physics scenarios will be reachable and studied at accelerators it will be possible cross-check theoretical models and predict effects measurable in cosmic ray astroparticle physics.

\section{Acknowledgements}
We would like to thank the PAMELA Collaboration for providing some of the 
information included in this paper and V. Formato for helping in the
compilation  
of this paper.






\bibliographystyle{model1-num-names}
\bibliography{ReviewCRspace}







\end{document}